\documentclass[aps,prl,reprint,preprintnumbers,groupedaddress,nofootinbib,floatfix,
  nolongbibliography]{revtex4-2}
\usepackage{amsmath,amssymb,graphicx,xcolor,booktabs}
\usepackage[colorlinks=true,citecolor=black,linkcolor=black,urlcolor=black]{hyperref}
\usepackage{xspace,orcidlink}
\newcommand{\neff}{\ensuremath{N_{\rm eff}}\xspace}
\newcommand{\yp}{\ensuremath{Y_{\rm P}}\xspace}
\providecommand{\dh}{}
\renewcommand{\dh}{\ensuremath{\mathrm{D/H}}\xspace}
\newcommand{\li}{\ensuremath{{}^{7}\mathrm{Li/H}}\xspace}
\newcommand{\hethree}{\ensuremath{{}^{3}\mathrm{He/H}}\xspace}

\newcommand{\tpeak}{\ensuremath{\tau_{\rm peak}}\xspace}
\newcommand{\nudsmc}{\ensuremath{\nu}\textsc{Dsmc}\xspace}
\newcommand{\MeV}{\ensuremath{\mathrm{MeV}}\xspace}
\newcommand{\keV}{\ensuremath{\mathrm{keV}}\xspace}
\definecolor{authornote}{rgb}{0.55,0.12,0.26}
\newif\ifauthornotes
\authornotesfalse

\makeatletter
\def\paragraph{\@startsection{paragraph}{4}{\parindent}{\z@}{-1em}{\normalfont\normalsize\bfseries\itshape}}
\makeatother
\begin{document}
\preprint{CERN-TH-2026-305}
\title{The cosmological lithium problem and how decaying relics still solve it}
\author{Maksym~Ovchynnikov\,\orcidlink{0000-0001-7002-5201}}
\email{maksym.ovchynnikov@cern.ch}
\affiliation{Theoretical Physics Department, CERN, 1 Esplanade des Particules, CH-1211 Geneva 23, Switzerland}
\hypersetup{pdftitle={The cosmological lithium problem and how decaying relics still solve it},pdfauthor={Maksym Ovchynnikov}}

\begin{abstract}
Lithium-7 in old stars is about three times below the prediction of standard big-bang nucleosynthesis, and deuterium lies slightly above it.
Precise measurements of deuterium, helium-4, and the effective number of neutrino species now disfavor most previously proposed particle solutions.
We find that relics that decay or evaporate with lifetimes of a few hours, mostly into electron antineutrinos of about a hundred MeV and into electrons or photons, still lower lithium-7 while respecting these measurements.
Such relics include gauge bosons of $L_e-L_\tau$ or $B-3L_e$ and heavy neutral leptons with a dipole coupling or mixing with $\nu_\tau$, with masses of a few hundred MeV, and populations of evaporating Q-balls that carry lepton number.
The decaying particles substantially alleviate the lithium problem, while Q-ball populations reach the $2\sigma$ range of lithium-7 adopted by the Particle Data Group.
These relics can shift the effective number of neutrino species by up to about $2.5$ times the precision expected from the Simons Observatory, and a future Galactic supernova can test the particles.
\end{abstract}
\maketitle
\paragraph{Introduction.}
Big-bang nucleosynthesis (BBN) and the cosmic microwave background (CMB) probe the Universe at keV to MeV temperatures through the primordial abundances of light nuclei and the effective number of relativistic species, \neff.
Since 1982, old, metal-poor halo stars have shown nearly the same lithium-7 abundance over a wide metallicity range, about a third of the predicted value~\cite{Spite:1982dd,Sbordone:2010zi,Charbonnel:2026lip}, as does interstellar gas in the Small Magellanic Cloud~\cite{Molaro:2024wxa}.
This uniformity suggests a deficit already present before the stars formed.
Surface depletion may explain it~\cite{Borisov:2024lid}, but its size depends on poorly known internal mixing~\cite{Charbonnel:2026lip}, so the primordial lithium-7 abundance remains uncertain.

Many non-standard scenarios have been proposed to lower primordial lithium-7, often with decaying or charged weak-scale particles such as superpartners~\cite{Fields:2011zzb,Jedamzik:2009uy,Pospelov:2010hj,Mathews:2019hbi,Jedamzik:2004er,Ellis:2005ii,Jedamzik:2005dh,Cumberbatch:2007me,Bailly:2008yy,Kohri:2008cf,Cyburt:2010vz,Cyburt:2013fda,Pospelov:2010cw,AlbornozVasquez:2012emy,Olive:2012xf,Coc:2013eha,Coc:2014gia,Kusakabe:2014ola,Ganguly:2026uzd,Wang:2004aq,Ichikawa:2004pb,Poulin:2015woa,Goudelis:2015wpa,Ishida:2014wqa,Salvati:2016jng,Alcaniz:2019kah,Kawasaki:2020qxm,Kohri:2006cn,Bird:2007ge,Jittoh:2007fr,Jedamzik:2007cp,Kusakabe:2007fu,Kawasaki:2010yh,Bailly:2010hh,Cyburt:2012kp,Kusakabe:2014moa,Coc:2006sx,Berengut:2009js,Clara:2020efx,Franchino-Vinas:2021nsf,Fung:2021wbz,Erken:2011vv,Yamazaki:2014fja,Luo:2018nth,Cuesta:2023awo,Larena:2005tu,Hou:2017uap,Talukdar:2024qny,Flambaum:2018ohm,Koren:2022axd,Arbuzova:2025ghi,Dent:2007zu,Dmitriev:2003qq,Flambaum:2008ps,Cheoun:2011yn,Landau:2004rj,Mori:2019cfo,Pospelov:2006sc,Yamazaki:2017uvc,Cyburt:2006uv,Kaplinghat:2006qr,Kusakabe:2010cb,Kusakabe:2011zz,Anchordoqui:2020djl,Sato:2016len,Jittoh:2011ni,Kusakabe:2017brd,Makki:2019zem,Malaney:1993ah,Mathews:2004kc,Nakamura:2017qtu,Lu:2022aus,Pospelov:2012pa}.
Many inject neutrons, which convert $^7$Be into $^7$Li that protons then destroy, but also produce deuterium.
Most were tested when \dh, \yp, and \neff had several times larger uncertainties.
Since \dh has been measured to about a percent~\cite{Cooke:2013cba,Cooke:2017cwo,Kislitsyn:2024dh,Guarneri:2024dh}, the review of the Particle Data Group concludes that particle injection changes $^7$Li far less than needed~\cite{ParticleDataGroup:2026bbn}.
Other proposals reduce lithium selectively or modify nucleosynthesis more broadly; with the recent measurements of \yp~\cite{Aver:2026dxv} and \neff~\cite{SPT-3G:2025bzu,Tristram:2025you} and the bounds on decaying relics~\cite{Kawasaki:2017bqm,Depta:2020zbh,EscuderoAbenza:2025tsi,Bianco:2026ibb}, their realizations would be constrained as well.
Finding realizations that respect all present measurements is therefore nontrivial.

Moreover, with the deuterium-burning rates of Refs.~\cite{Pitrou:2018cgg,Pitrou:2020etk}, standard BBN predicts \dh about $2\sigma$ below the measurements~\cite{EscuderoAbenza:2026dqx,Poulin:2026ltf}, a deficit that other scenarios address on its own~\cite{Poulin:2026ltf,Elahi:2026ldn,Cuesta:2026hlk,EscuderoAbenza:2025tsi,DiBari:2024jkj,Masip:2024vow,DiBari:2023upq,Cuesta:2023awo,Tian:2022bwz,Deal:2021kjs}.
Resolving both tensions requires lowering $^7$Li about threefold while raising \dh by only a few percent and keeping \yp and \neff within their present intervals.
New measurements of nuclear rates and stellar lithium will establish whether these tensions persist.

\begin{figure}[t!]
\centering
\includegraphics[width=0.92\columnwidth]{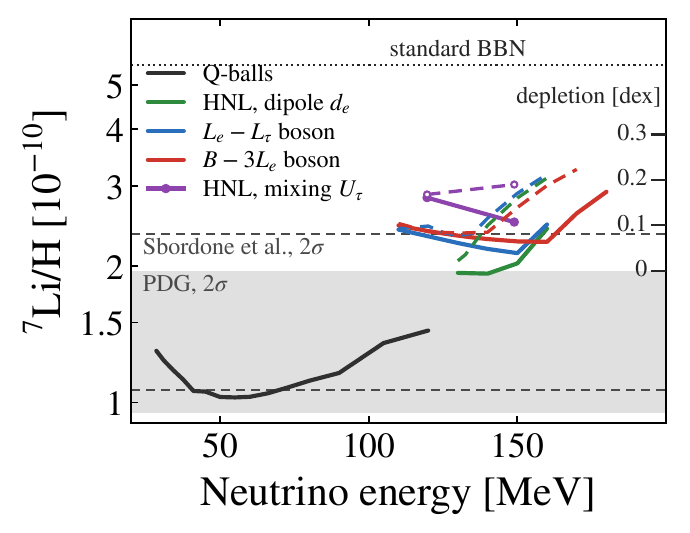}\\[-2pt]
\includegraphics[width=0.92\columnwidth]{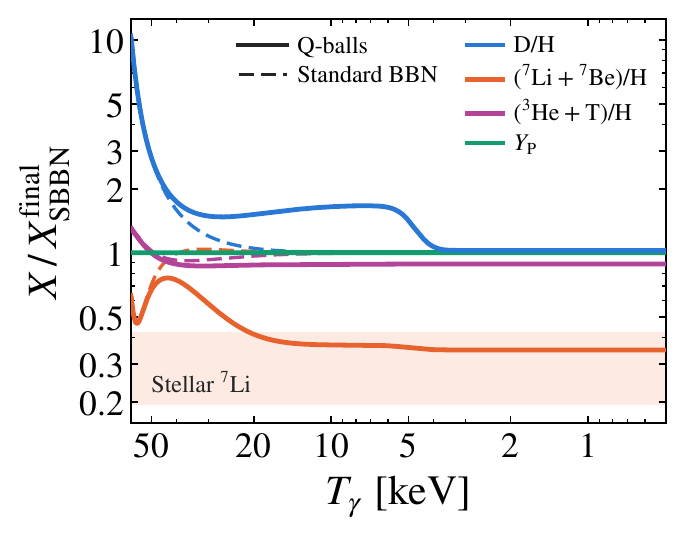}
\caption{Upper panel: lowest \li found for each relic with \dh, \yp, \hethree, and \neff within Eqs.~\eqref{eq:abundances} and \eqref{eq:lithium}, versus the energy of the emitted neutrinos, at the most probable charge for Q-balls. Heavy-neutral-lepton masses excluded by SN~1987A~\cite{Brdar:2023tmi,Chauhan:2025sterileSN} are not shown. Solid and dashed colored curves: cross sections for $^4$He breakup from the two shell-model Hamiltonians of Ref.~\cite{Yoshida:2008zb}, WBP and SPSDMK, respectively. Gray band and horizontal lines: stellar $^7$Li at $2\sigma$~\cite{ParticleDataGroup:2026bbn,Sbordone:2010zi}, with the depletion needed to enter the range of Ref.~\cite{ParticleDataGroup:2026bbn} at the right. Lower panel: abundances divided by their final values in standard BBN (SBBN), for SBBN (dashed) and Q-ball population 1 of Table~\ref{tab:solutions} (solid); orange band: stellar $^7$Li~\cite{Sbordone:2010zi,Charbonnel:2026lip}.}
\label{fig:qball-abundances}
\end{figure}

In this paper, we identify unstable relics that address both tensions.
They bring \dh into its observed range and substantially lower $^7$Li, while helium, $^3$He, and \neff stay within their bounds.
The successful sources combine antineutrinos of about $100\,\MeV$, which convert protons into neutrons while $^7$Be forms, with electrons or photons carrying about half of the emitted power, whose later cascades destroy the extra deuterium.
Gauge bosons of $L_e-L_\tau$ or $B-3L_e$ and heavy neutral leptons with a dipole coupling or mixing with $\nu_\tau$ produce this emission at masses of $240$ to $350\,\MeV$ and lower $^7$Li two- to threefold (Table~\ref{tab:solutions}, Fig.~\ref{fig:qball-abundances}).
Populations of leptonic Q-balls, nontopological solitons that carry lepton number and evaporate through their surface, also resolve both tensions and take $^7$Li further, into the $2\sigma$ range of Ref.~\cite{ParticleDataGroup:2026bbn} (Fig.~\ref{fig:qball-abundances}), because the smallest release harder electrons and neutrinos early and the largest softer ones later.

The Supplemental Material (SuM) reviews previous solutions and details the calculation, the particle models, and the Q-ball populations.

\paragraph{Observational constraints.}
We adopt $10^5\dh=2.508\pm0.029$ and $\yp=0.2458\pm0.0013$ at $1\sigma$~\cite{ParticleDataGroup:2026bbn,Aver:2026dxv}, and bound \hethree, counting surviving tritium, at $2\sigma$ above the upper limit on primordial $^3$He/H inferred from the Orion nebula~\cite{Cooke:2022cvb}.
For \neff, we take the range spanned at $2\sigma$ by recent CMB analyses treating foregrounds and helium differently~\cite{SPT-3G:2025bzu,EscuderoAbenza:2026dqx,Tristram:2025you}.
These give the intervals
\begin{equation}
 \begin{gathered}
 2.450\leq10^5\dh\leq2.566,\quad
 \hethree\leq1.45\times10^{-5},\\
 0.2432\leq\yp\leq0.2484,\quad
 2.57\leq\neff\leq3.291 .
 \end{gathered}
 \label{eq:abundances}
\end{equation}
We call a scenario allowed when it satisfies Eq.~\eqref{eq:abundances}.

For \li, the Particle Data Group adopts $(1.45\pm0.25)\times10^{-10}$ from metal-poor stars, a lower bound on the primordial value if they destroyed part of their lithium~\cite{ParticleDataGroup:2026bbn}.
A scenario with more \li than $1.95\times10^{-10}$, the upper end of this range at $2\sigma$, needs stellar depletion of the excess, which Fig.~\ref{fig:qball-abundances} shows in dex.
The scatter of the lithium plateau~\cite{Sbordone:2010zi} sets a wider $2\sigma$ band, and we call lithium solutions the allowed scenarios with
\begin{equation}
 1.064\leq10^{10}\,\li\leq2.350,\quad
 2.69\leq\neff\leq3.161,
 \label{eq:lithium}
\end{equation}
where the \neff interval is the union of the $1\sigma$ intervals of the same CMB analyses.
Spectra of three metal-poor stars give upper limits $^6\mathrm{Li}/{}^7\mathrm{Li}=0.006$ to $0.017$ at $2\sigma$~\cite{Wang:2021urb}; since stars destroy $^6$Li more readily than $^7$Li and the inferred $^6$Li depends on the line profile~\cite{Charbonnel:2026lip,Wang:2021urb}, we use these limits for comparison only.

\paragraph{Routes to lower lithium-7.}
In standard BBN, most primordial $^7$Li forms as $^7$Be, which captures an electron long after nucleosynthesis, so lowering $^7$Li means forming less $^7$Be or destroying more of it~\cite{Jedamzik:2004er}.
The proposed mechanisms follow four main routes; below, we discuss each and whether its existing solutions satisfy Eqs.~\eqref{eq:abundances} and \eqref{eq:lithium}.
\begin{itemize}
\item \emph{Extra neutrons.}
Until about $3\times10^3\,\mathrm{s}$, neutrons convert $^7$Be into $^7$Li through $^7\mathrm{Be}(n,p){}^7\mathrm{Li}$, and protons destroy this $^7$Li through ${}^7\mathrm{Li}(p,\alpha){}^4\mathrm{He}$.
Neutrons injected by decaying particles or made from protons by energetic electron antineutrinos mostly end in deuterium, so they reach the stellar $^7$Li only with \dh well above its measurement~\cite{Coc:2014gia}.
Ref.~\cite{Goudelis:2015wpa} avoids this with light particles that free neutrons from deuterons, to which the neutrons return; our analysis of solar data, kaon decays, and the decays of these particles at larger masses excludes the proposed benchmark and the minimal completions that we examined~\cite{Borexino:2012guz,BOREXINO:2025dbp,NA62:2025upx}.
\item \emph{Selective destruction.}
Existing studies considered photons between the breakup thresholds of $^7$Be and D, $1.59$ and $2.22\,\MeV$, which destroy $^7$Be and spare deuterium; such photons escape pair production on background photons only below about $7\,\keV$, so their sources must live about $10^4$ to $10^6\,\mathrm{s}$~\cite{Poulin:2015woa,Kawasaki:2020qxm}.
They also considered electrically charged relics, which bind $^7$Be below about $30\,\keV$ and catalyze its destruction by protons~\cite{Pospelov:2006sc,Kusakabe:2014moa}, and cosmic strings, which destroy lithium after nucleosynthesis~\cite{Koren:2022axd}.
By themselves, these mechanisms leave \dh at its standard value.
Supernova data and CMB spectral distortions constrain the minimal particles that emit the photons~\cite{Jaeckel:2017tud,Muller:2023vjm,Caputo:2022mah,Balazs:2022tjl,EscuderoAbenza:2025tsi}.
\item \emph{Changed conditions of nucleosynthesis.}
Other studies changed the conditions of nucleosynthesis through primordial magnetic fields~\cite{Yamazaki:2014fja,Luo:2018nth,Lu:2022aus}, varying fundamental constants~\cite{Coc:2006sx,Dent:2007zu,Berengut:2009js,Mori:2019cfo,Clara:2020efx,Franchino-Vinas:2021nsf}, non-extensive statistics~\cite{Hou:2017uap}, or modified gravity~\cite{Larena:2005tu,Talukdar:2024qny}, which alter all abundances at once; several of the solutions they report conflict with the present measurements, Eqs.~\eqref{eq:abundances} and \eqref{eq:lithium}.
\item \emph{Neutrons followed by photodisintegration.}
Ref.~\cite{Pospelov:2010cw} proposed that, once the injected neutrons have destroyed $^7$Be, electromagnetic cascades remove the deuterium that these neutrons made, and realized this with a particle that decays into muon pairs.\footnote{Ref.~\cite{Ganguly:2026uzd} lowers $^7$Li and restores \dh with majorons and axion-like particles whose masses, lifetimes, and abundances are chosen together.}
Photons above $E_c\simeq m_e^2/(22T_\gamma)$ turn into electron-positron pairs on background photons, so the cascade photons that reach nuclei lie below $E_c$~\cite{Depta:2020mhj}; they break up $^7$Be below about $7\,\keV$ and deuterium from about $5\,\keV$ on~\cite{Kawasaki:2020qxm}.
This realization brings $^7$Li into the stellar band only with \neff and \yp above their present limits, since neutrinos carry most of the energy of muon decays and only the $\bar\nu_e$ among them convert protons into neutrons.
\end{itemize}
This last route needs only one source with ordinary decay products.
Its energetic electron antineutrinos convert protons into neutrons directly, without the hadronic showers that dissociate helium, and its electrons or photons later destroy the extra deuterium.
The failure of the muon pairs motivates harder electron antineutrinos; we determine which emission meets all present bounds and which sources produce it.

\paragraph{Evolution of relics and cosmic plasma.}
We evolve the neutrino distributions with \nudsmc~\cite{Ovchynnikov:2024rfu,Ovchynnikov:2024xyd,Ihnatenko:2025kew,Akita:2026xwu}.
We follow neutrinos and antineutrinos separately and add the emission of the decaying particles or Q-balls, muon propagation and decay, and the nuclear reactions of energetic neutrinos and breakup fragments.
The electromagnetic cascades of the emitted electrons and photons and the resulting photodisintegration rates come from the \textsc{Acropolis} module~\cite{Depta:2020mhj}.
The nuclear network follows Ref.~\cite{Akita:2026xwu}, extended with these rates and the nonthermal reactions, and photon heating and the remaining mass of the relics set the baryon density and expansion (SuM Sec.~\ref{sec:transport}).

We use the nuclear rates of Ref.~\cite{Pitrou:2018cgg} and fix the final baryon-to-photon ratio in every calculation to the CMB value, $\eta_B=6.13\times10^{-10}$~\cite{Planck:2018vyg}; standard BBN then gives $10^5\dh\simeq2.44$, as in Ref.~\cite{Pitrou:2020etk}.
A new evaluation of these rates from the measured cross sections~\cite{Launders:2026ciu} agrees with them and suggests that Ref.~\cite{Pisanti:2020efz} finds a higher \dh from its fits of these data.

\paragraph{Required decay products and lifetimes.}
Consider a relic $X$ of mass $m_X$ that decays with lifetime $\tau_X$ into pairs of particles, each with energy $m_X/2$.
Its abundance enters through the largest energy density of $X$ relative to radiation after neutrino decoupling,
\begin{equation}
 f=\max_{t>t_{\rm dec}}\frac{\rho_X(t)}{\rho_\gamma(t)+\rho_\nu(t)},\qquad \rho_X\propto a^{-3}e^{-t/\tau_X},
 \label{eq:f}
\end{equation}
where the scale factor $a$ and the photon and neutrino energy densities $\rho_\gamma$ and $\rho_\nu$ follow standard BBN; for Q-balls, $\rho_X$ is the energy density of the population, which also falls as the solitons lose mass, Eq.~\eqref{eq:evolution}.
We treat $f$ as a free parameter.
The measurements restrict $\tau_X$, the decay products, and their energies.
Its lifetime must lie between about $10^2$ and $10^6\,\mathrm{s}$.
Neutrons made earlier, before $^7$Be forms, end mostly in helium and raise \yp, whereas after about $10^6\,\mathrm{s}$ the electromagnetic cascades of the emitted electrons and photons break up $^4$He, overproduce $^3$He relative to deuterium, and distort the CMB spectrum~\cite{Kawasaki:2017bqm,Kawasaki:2020qxm}.
We take $X$ nonrelativistic from the freeze-out of neutrons and protons on, near $1\,\mathrm{s}$.

Decays with a sizable hadronic branching at $10^2$ to $10^4\,\mathrm{s}$ dissociate $^4$He, about $3\times10^3$ times more abundant than deuterium, and so raise \dh more than cascades can later reduce it~\cite{Kawasaki:2017bqm,Bianco:2026ibb}.
We therefore consider decays without a sizable hadronic branching, into neutrinos, photons, and the charged leptons $e^\pm$ and $\mu^\pm$.

For $m_X$ of a few MeV, lowering $^7$Li requires so many neutrinos that \neff exceeds its bound.
Decays of $X$ into electrons or photons heat the photons and could lower \neff again, but with the baryon-to-photon ratio fixed by the CMB, this heating after nucleosynthesis means a higher ratio during it, which lowers \dh and raises $^7$Li and \yp.

Destroying enough $^7$Li without excess \neff requires a larger $m_X$, and hence more energetic neutrinos, because per neutrino the rate of proton conversion grows as $E_\nu^2$, whereas the contribution to \neff grows only as $E_\nu$.
For antineutrino energies $E_\nu=m_X/2$ near $100\,\MeV$, the antineutrinos therefore convert enough protons to bring $^7$Li into the stellar band before \yp or \neff reach their limits (SuM Sec.~\ref{sec:single-hard}).
Their neutrons also raise \dh, so the electromagnetic cascades of the electrons and photons that $X$ emits must then destroy this extra deuterium.
Since these cascades reach the deuterium threshold near $5\times10^4\,\mathrm{s}$, $\tau_X$ is then near $10^4\,\mathrm{s}$, and only about a quarter of the decays precede $3\times10^3\,\mathrm{s}$.

The same neutrinos and antineutrinos also break up $^4$He into tritium and $^3$He, whose fusion with ambient helium produces $^6$Li.
The number of these fragments per converted proton grows steeply with the neutrino energy $E_\nu=m_X/2$~\cite{Yoshida:2008zb}, so lowering $^7$Li with harder neutrinos makes more $^3$He.
For equal numbers of electron neutrinos and antineutrinos, \hethree exceeds its bound once $E_\nu$ exceeds about $150\,\MeV$, that is, for $m_X$ above about $300\,\MeV$.
In our calculations, harder emission does not escape this bound, since every charged-current capture on $^4$He breaks it and, above about $200\,\MeV$, harder antineutrinos convert no more protons per unit of energy~\cite{LlewellynSmith:1971uhs,Formaggio:2012cpf}.

Therefore, for equal numbers of neutrinos and antineutrinos, the measurements leave $m_X$ of a few hundred MeV, up to about $320\,\MeV$, with the lower end depending on the decay products (upper panel of Fig.~\ref{fig:qball-abundances}), and about half of the energy in electrons or photons and the rest mostly in electron neutrinos and antineutrinos.\footnote{With more antineutrinos than neutrinos, solutions extend to antineutrino energies from about $75\,\MeV$ to at least $230\,\MeV$; an asymmetric fermion decaying through a scalar realizes them if an earlier process sets its asymmetry.}

\paragraph{Particle models.}
Three particles with one coupling each produce this emission (Table~\ref{tab:solutions}): the gauge bosons $Z'$ of $L_e-L_\tau$~\cite{He:1990pn,Foot:1990mn} and of $B-3L_e$~\cite{Ma:1997nq,Chang:2000xy}, and a heavy neutral lepton (HNL) $N$~\cite{Magill:2018jla}, with the couplings
\begin{equation}
 \mathcal L\supset g\,Z'_\mu\sum_f Q_f\,\bar f\gamma^\mu f+\left(d_e\,\bar\nu_{eL}\sigma_{\mu\nu}N F^{\mu\nu}+\mathrm{h.c.}\right),
 \label{eq:couplings}
\end{equation}
where $f$ runs over the Standard Model fermions, with charges $Q_f=1$ for $e$ and $\nu_e$ and $-1$ for $\tau$ and $\nu_\tau$ ($L_e-L_\tau$), or $1/3$ for quarks and $-3$ for $e$ and $\nu_e$ ($B-3L_e$), with left-handed neutrinos.
The bosons decay into $e^+e^-$, $\nu_e\bar\nu_e$, and $\nu_\tau\bar\nu_\tau$ as $2:1:1$, or into $e^+e^-$ and $\nu_e\bar\nu_e$ as $2:1$.
The HNL yields $\nu_e\gamma$ and $\bar\nu_e\gamma$ in equal numbers if it is a Majorana fermion, or a Dirac fermion without asymmetry, and its photon pair-produces at once on background photons.

\begin{table*}[t!]
\centering
\footnotesize
\begin{tabular}{@{}l@{\hspace{4.5pt}}l@{\hspace{4.5pt}}c@{\hspace{4.5pt}}c@{\hspace{4.5pt}}c@{\hspace{4.5pt}}c@{\hspace{4.5pt}}c@{\hspace{4.5pt}}c@{\hspace{4.5pt}}c@{}}
\toprule
Solution & Parameters & $10^2f$ & \begin{tabular}[b]{@{}c@{}}$\tau_X$ or \tpeak\\$[10^3$\,s]\end{tabular} & \yp & $10^5\,{}^3\mathrm{He/H}$ & $10^{10}\,{}^7\mathrm{Li/H}$ & \neff & $10^2\,{}^6\mathrm{Li}/{}^7\mathrm{Li}$\\
\midrule
Gauge boson of $L_e-L_\tau$ & $m_X=260\,\MeV$, $g=7.5\times10^{-14}$ & $5.2$ & $8.5$ & $0.2481$ & $1.20$ & $2.32$ & $3.15$ & $6.1$ to $15$\\
Gauge boson of $B-3L_e$ & $m_X=260\,\MeV$, $g=3.0\times10^{-14}$ & $5.5$ & $8.1$ & $0.2484$ & $1.14$ & $2.29$ & $2.95$ & $5.2$ to $13$\\
HNL, dipole $d_e$ & $m_X=265\,\MeV$, $d_e=1.6\times10^{-13}\,\mathrm{GeV}^{-1}$ & $4.4$ & $8.9$ & $0.2479$ & $1.21$ & $2.18$ & $3.14$ & $7.1$ to $17$\\
\midrule
Q-balls, 1\,$\bigstar$ & $\omega_{\rm peak}=41\,\MeV$, $b=0.75$, $\epsilon=-1$ & $1.7$ & $12.4$ & $0.2476$ & $0.922$ & $1.94$ & $3.09$ & $0.57$ to $1.19$\\
Q-balls, 2\,$\blacklozenge$ & $\omega_{\rm peak}=55\,\MeV$, $b=0.9$, $\epsilon=-1$ & $1.4$ & $12.1$ & $0.2476$ & $0.950$ & $2.18$ & $3.04$ & $0.76$ to $1.54$\\
Q-balls, 3\,$\blacktriangle$ & $\omega_{\rm peak}=28.5\,\MeV$, $b=1$, $\epsilon=-1$ & $3.3$ & $10.4$ & $0.2482$ & $0.870$ & $2.00$ & $2.97$ & $0.17$ to $0.39$\\
Q-balls, 4 & $\omega_{\rm peak}=31\,\MeV$, $b=0.7$, $\epsilon=-0.5$ & $3.0$ & $11.7$ & $0.2479$ & $0.878$ & $1.52$ & $3.15$ & $0.53$ to $1.18$\\
\midrule
SBBN & & & & $0.2471$ & $1.040$ & $5.53$ & $3.044$ & \\
\bottomrule
\end{tabular}
\caption{Lithium solutions, Eq.~\eqref{eq:lithium}, of particles with the couplings of Eq.~\eqref{eq:couplings} and of Q-ball populations, at the lifetime $\tau_X$, or \tpeak, at which $10^5\dh=2.508$, and standard BBN (SBBN, $10^5\dh=2.443$).
$f$: Eq.~\eqref{eq:f}; $d_e$ for a Majorana HNL; means of four independent simulations.
${}^3\mathrm{He/H}$ includes tritium; $^6\mathrm{Li}/{}^7\mathrm{Li}$ spans two recoil prescriptions for helium fragments.}
\label{tab:solutions}
\end{table*}

\paragraph{Leptonic Q-balls.}
For one decaying species, the $\tau_X$ that the measured \dh requires also fixes the fraction of decays that occur while neutrons still lower $^7$Li.
Relics with a range of lifetimes change this fraction and tie the energy of the emitted particles to the time of emission; populations of leptonic Q-balls are such relics.
Each Q-ball is a localized configuration of a complex scalar field that has the lowest energy at fixed conserved charge $Q$, with mass $M_Q$~\cite{Coleman:1985ki}.
Q-balls form when a homogeneous condensate of this field in the early Universe fragments~\cite{Affleck:1984fy,Dine:1995kz}, so they have a nonthermal abundance and inherit the net charge of the condensate.
Earlier proposals lowered $^7$Li with Q-balls that release baryons~\cite{Jedamzik:2004er,Wang:2004aq,Ichikawa:2004pb}, which now raise \dh above its measured interval.

We consider an electrically neutral field that carries lepton number.
To describe a population of its Q-balls, we need the mass and charge-loss rate of each soliton, the distribution of initial charges, and the composition of the emission.
These Q-balls evaporate at their surface, where pairs of quanta turn into pairs of electrons or neutrinos~\cite{Cohen:1986ct}, so each Q-ball emits one electron or neutrino per unit of charge, carrying on average the energy $\omega_Q=dM_Q/dQ$.
When the scalar potential is nearly constant, at a value $V_0$, over the field values inside a Q-ball, its mass is about $V_0$ times its volume, and its radius grows as $Q^{1/4}$.
Then~\cite{Kasuya:2025nix}
\begin{equation}
 M_Q=\frac{4}{3}C Q^{3/4},\qquad
 \omega_Q=\frac{dM_Q}{dQ}=C Q^{-1/4} ,
 \label{eq:mass}
\end{equation}
where $C$ combines $V_0^{1/4}$ with a numerical factor from the field profile.

Pauli blocking of the outgoing fermions limits the evaporation rate per unit area to a value proportional to $\omega_Q^3$~\cite{Cohen:1986ct}, so a Q-ball that emits at this maximal rate into a fixed set of channels loses charge as $\dot Q=-AQ^{-1/4}$, where $A=\pi C/3$~\cite{Kawasaki:2012gk,Kasuya:2025nix}.
Solving this equation for an initial charge $Q_0$, we get
\begin{equation}
 \begin{gathered}
 \tau_Q=\frac{4Q_0^{5/4}}{5A},\qquad
 \frac{Q(t)}{Q_0}=\left(1-\frac{t}{\tau_Q}\right)^{4/5},\\
 \frac{M_Q(t)}{M_Q(0)}=\left(1-\frac{t}{\tau_Q}\right)^{3/5}.
 \end{gathered}
 \label{eq:evolution}
\end{equation}
Here, $t$ is the age of a soliton, $0\leq t<\tau_Q$.
Large Q-balls thus live longer and emit less energetic electrons and neutrinos, whose energy grows as each soliton shrinks and loses its mass in a finite time.

We use the distribution of initial energy density $\rho_{Q,0}$ found in numerical simulations of fragmentation~\cite{Kasuya:2025nix}.
In terms of $x=Q_0/Q_{\rm peak}$, with $Q_{\rm peak}$ the charge at its maximum, it is
\begin{equation}
 \begin{gathered}
 \frac{d\rho_{Q,0}}{d\ln x}=\rho_{Q,0}W(x),\qquad
 \int W(x)\,d\ln x=1,\\
 W(x)=\mathcal N x^{1/2}e^{-x^2/4} .
 \end{gathered}
 \label{eq:distribution}
\end{equation}
Solitons with this charge define the \emph{peak lifetime} $\tpeak\equiv\tau_Q(Q_{\rm peak})$ and the characteristic injection energy $\omega_{\rm peak}\equiv\omega_Q(Q_{\rm peak})$ at age $t=0$; they play the roles of $\tau_X$ and $m_X/2$ of a decaying relic.
Thus, $\tau_Q(x)=\tpeak x^{5/4}$ and $\omega_Q(x,0)=\omega_{\rm peak}x^{-1/4}$.

As in Ref.~\cite{Kasuya:2025nix}, the scalar field combines three fields with equal amplitudes, whose quanta yield $\nu_\mu$, $e^-$ or $\nu_e$, and $e^+$ or $\bar\nu_e$, respectively.
One third of the emitted particles are therefore $\nu_\mu$, and the others are electrons and positrons in equal numbers or $\nu_e$ and $\bar\nu_e$, in the ratio set by the charged fraction $b$,
\begin{equation}
 N_{\nu_\mu}:N_{e^\pm}:N_{\nu_e}=1:2b:2(1-b),
 \label{eq:composition}
\end{equation}
where $N_{\nu_\mu}$, $N_{e^\pm}$, and $N_{\nu_e}$ count both particles and antiparticles, with $b=0$ for $\omega_Q$ below the electron mass.
We call these electron populations.
Q-balls of antiquanta emit the charge-conjugate particles.
A population contains both kinds, with energy fractions $(1+\epsilon)/2$ of Q-balls of quanta and $(1-\epsilon)/2$ of antiquanta; we mainly study $\epsilon=-1$, for which the excess particles are $\bar\nu_\mu$, which oscillations partly turn into $\bar\nu_e$ that convert protons into neutrons.

We consider populations created long before nucleosynthesis and extend Eq.~\eqref{eq:distribution} to charges below those simulated, which evaporate earliest.
Since $C$ and $Q_{\rm peak}$ enter only through $\omega_{\rm peak}$ and \tpeak, and fragmentation fixes the shape of $W$, a population has five parameters: $\omega_{\rm peak}$, \tpeak, $b$, $\epsilon$, and $R_Q$.
Among them, $R_Q=\rho_Q/\rho_{\rm rad}$ at $T_\gamma=3\,\MeV$, where the calculation starts, sets the abundance.
Together with \tpeak, it fixes the energy fraction $f$ of Eq.~\eqref{eq:f}.
A microscopic model would fix these parameters (SuM Sec.~\ref{sec:embedding}).

\paragraph{Lifetimes and abundances of the solutions.}
For particles and Q-ball populations alike, the effect of the cascades on \dh grows steeply with the lifetime (lower panel of Fig.~\ref{fig:qball-abundances}), so the measured \dh fixes the lifetime within a few percent, near $9\times10^3\,\mathrm{s}$ for the particles and $(1$ to $2.1)\times10^4\,\mathrm{s}$ for the populations (Table~\ref{tab:solutions}).
The stellar $^7$Li band and the limits on \yp, \neff, and \hethree then confine $f$ to a narrow interval.
With \dh at the lower end of its measured interval, next to its standard value, $^7$Li changes by half a percent, so the solutions do not rely on the deuterium tension.

At these lifetimes, the couplings of the three particles evade the bounds from supernovae with a cold core~\cite{Jaeckel:2017tud,Caputo:2022mah,Brdar:2023tmi,Chauhan:2025xqr}, neutron-star mergers~\cite{Diamond:2023cto}, and laboratory searches~\cite{Bauer:2018onh,Magill:2018jla} (SuM Sec.~\ref{sec:supernovae}); a hot one, or a detector response to photons outside the analyzed band, may exclude some HNL masses and the $B-3L_e$ boson.
All three bring \yp close to the upper limit of Eq.~\eqref{eq:abundances} and give $^6\mathrm{Li}/{}^7\mathrm{Li}$ several times the stellar upper limits (Table~\ref{tab:solutions}).

One decaying species emits with a single lifetime, which \dh fixes near $10^4\,\mathrm{s}$, after ${}^7\mathrm{Li}(p,\alpha){}^4\mathrm{He}$ stops at about $3\times10^3\,\mathrm{s}$; most of its neutrons then turn $^7$Be into $^7$Li that survives, so it needs a larger abundance, which raises \yp.
In a Q-ball population, the small solitons release almost half of the energy, in antineutrinos of more than twice $\omega_{\rm peak}$, before nuclear burning ends, and the large ones later supply the softer electrons that destroy the extra deuterium, so it reduces $^7$Li further with a lower $f$ and \yp and, for $\omega_{\rm peak}\lesssim50\,\MeV$, less $^6$Li (Table~\ref{tab:solutions}, SuM Sec.~\ref{sec:emission}).
Before \yp or \hethree reaches its bound, Q-ball populations bring \li down to about $1.0\times10^{-10}$, half the minimum for the particles (Fig.~\ref{fig:qball-abundances}).

A larger $b$ moves energy from neutrinos to electrons and photons, so across lithium solutions \neff falls from the upper limit of Eq.~\eqref{eq:lithium}, reached for $b\lesssim0.65$, to about $2.96$ at $b=1$, where the upper limit on \yp stops it.
\yp ranges from its standard value, for $b\approx0.3$, to the bound of Eq.~\eqref{eq:abundances}, while $^6\mathrm{Li}/{}^7\mathrm{Li}$ and \hethree rise with $\omega_{\rm peak}$, \hethree from a deficit relative to standard BBN to an excess.

The solutions persist for lifetime-charge relations, emission spectra, and charge distributions that keep energetic antineutrinos early and enough electromagnetic energy hours later (SuM Sec.~\ref{sec:variations}).

\paragraph{Conclusions.}
Unstable relics resolve the lithium and deuterium tensions when they emit antineutrinos of about $100\,\MeV$ while $^7$Be forms and electromagnetic energy hours later.
Gauge bosons of $L_e-L_\tau$ or $B-3L_e$ and heavy neutral leptons with a dipole coupling to $\nu_e$ and the photon or mixing with $\nu_\tau$ strongly alleviate the lithium problem in this way; to reach the $2\sigma$ range of primordial $^7$Li adopted by the Particle Data Group, they need slight depletion of lithium in stars.
The couplings that set the lifetimes of the particles also make them in supernovae and may be tested by future observations of a Galactic supernova.

Q-ball populations emit harder electrons and neutrinos early and softer ones later, so they need a smaller abundance than the particles, reach the range of the Particle Data Group without stellar depletion, and can form less $^6$Li.
Supernovae cannot produce them, so they remain a possible solution even if a Galactic supernova excludes the particles.
Their condensate can also yield the observed baryon asymmetry if a few times $10^{-5}$ of its lepton number escapes capture before sphalerons stop (SuM Sec.~\ref{sec:provide}).

For each particle mass or Q-ball population, the measured \dh and $^7$Li restrict the lifetime and the abundance to narrow ranges, so that its \neff, \yp, and $^6\mathrm{Li}/{}^7\mathrm{Li}$ are predictions.
These relics shift \neff by up to about $2.5$ times the precision expected from the Simons Observatory~\cite{SimonsObservatory:2025wwn}, a change that lithium depletion in stars does not produce.

\section*{Acknowledgements}
The author thanks Matei Climescu and Maxim Pospelov for reading the manuscript, and Maxim Pospelov also for discussions on particle solutions of the lithium problem.
The author received support from the European Union's Horizon Europe research and innovation programme under the Marie Sk{\l}odowska-Curie grant agreement No.~101204216.
The AI assistants Claude Opus 5.5 (Anthropic) and Astra (OpenAI) were used for code development and numerical cross-checks.
The author takes full responsibility for the calculations, their interpretation, and the manuscript.

\bibliography{qballs}
\onecolumngrid
\clearpage
\begin{center}\textbf{\large Supplemental Material}\end{center}
\setcounter{equation}{0}
\setcounter{figure}{0}
\setcounter{table}{0}
\setcounter{section}{0}
\setcounter{secnumdepth}{1}
\renewcommand{\theequation}{S\arabic{equation}}
\renewcommand{\thefigure}{S\arabic{figure}}
\renewcommand{\thetable}{S\arabic{table}}
\renewcommand{\thesection}{\Alph{section}}
\renewcommand{\thesubsection}{\thesection.\arabic{subsection}}
\makeatletter\renewcommand{\p@subsection}{}\makeatother
\setcounter{secnumdepth}{2}
\renewcommand{\theHequation}{S\arabic{equation}}
\renewcommand{\theHfigure}{S\arabic{figure}}
\renewcommand{\theHtable}{S\arabic{table}}
\renewcommand{\theHsection}{S\arabic{section}}
This Supplemental Material describes the coupled neutrino, electromagnetic, and nuclear evolution and its numerical accuracy (Sec.~\ref{sec:transport}).
It compares earlier proposals and decaying particles, with their observational constraints, with the evaporation of Q-balls (Secs.~\ref{sec:single} to \ref{sec:emission}).
The final sections give the allowed Q-ball populations and their dependence on the emission and the charge distribution (Sec.~\ref{sec:lifetimes}), and the requirements for a supersymmetric Affleck-Dine realization (Sec.~\ref{sec:embedding}).

We call a decaying particle or a Q-ball population allowed when its deuterium-to-hydrogen ratio \dh, helium mass fraction \yp, \hethree including tritium, and effective number of relativistic species \neff satisfy
\begin{equation}
 2.450\leq10^5\dh\leq2.566,\quad
 0.2432\leq\yp\leq0.2484,\qquad
 \hethree\leq1.45\times10^{-5},\quad
 2.57\leq\neff\leq3.291,
 \label{eq:S-abundances}
\end{equation}
from the measured \dh and \yp at $2\sigma$~\cite{ParticleDataGroup:2026bbn,Aver:2026dxv}, the primordial $^3$He/H inferred from the Orion nebula~\cite{Cooke:2022cvb}, and CMB analyses~\cite{SPT-3G:2025bzu,EscuderoAbenza:2026dqx,Tristram:2025you}.
Lithium solutions are allowed particles or populations with
\begin{equation}
 1.064\leq10^{10}\,\li\leq2.350,\quad
 2.69\leq\neff\leq3.161,
 \label{eq:S-lithium}
\end{equation}
the stellar $^7$Li band~\cite{Sbordone:2010zi} and the union of the $1\sigma$ \neff intervals of the same CMB analyses.
The main text also compares with the range $(1.45\pm0.25)\times10^{-10}$ of Ref.~\cite{ParticleDataGroup:2026bbn}.
The Q-ball populations 1 to 3 of Table~\ref{tab:solutions} of the main text, three electron populations (Sec.~\ref{sec:source}) with $\epsilon=-1$ that are lithium solutions, serve as examples throughout (Table~\ref{tab:soft-points}).

\section{Neutrino transport, electromagnetic cascades, and nuclear evolution}\label{sec:transport}
In this section, we describe how we compute the light-element abundances and \neff.

Decaying particles and Q-ball populations change the light-element abundances through three coupled processes: their neutrinos convert nucleons and break up helium, their electrons, positrons, and photons produce electromagnetic cascades that break up nuclei, and both heat the plasma and change the expansion.
We follow the neutrinos with \nudsmc~\cite{Ovchynnikov:2024rfu,Ovchynnikov:2024xyd,Ihnatenko:2025kew,Akita:2026xwu}, the cascades with \textsc{Acropolis}~\cite{Depta:2020mhj}, and the nuclei with the network of Ref.~\cite{Akita:2026xwu}, extended with the photodisintegration rates and the nonthermal reactions.

\subsection{Neutrino transport}\label{sec:nu-transport}
\nudsmc represents neutrinos and antineutrinos by Monte Carlo particles.
At the initial photon temperature $T_0$ ($3\,\MeV$, or $5\,\MeV$ for $\tpeak\leq0.3\,\mathrm{s}$; Sec.~\ref{sec:source}), it samples $N_\nu$ particles from the thermal distributions of the three flavors; we call $N_\nu$ the number of simulated neutrinos.
Fictitious collision cells, which partition the homogeneous volume, only organize the sampling of collisions; their number adapts to a fixed target number of candidate collisions per cell and time step, above a set minimum, and Pauli blocking uses the occupations of the final states~\cite{Akita:2026xwu}.

For a decaying particle, \nudsmc injects the products at a rate proportional to $e^{-t/\tau}$ with a fixed spectrum.
For a Q-ball population, it injects the time-dependent emission of Sec.~\ref{sec:source}, Eq.~\eqref{eq:S-evolution}; the emitted neutrinos join the collision calculation as weighted Monte Carlo particles.

Ref.~\cite{Akita:2026xwu} evolves neutrinos and antineutrinos of each flavor as one population, which holds without a lepton asymmetry.
Populations with $\epsilon\neq0$, however, emit unequal numbers of neutrinos and antineutrinos, and the two convert nucleons in opposite directions: $\nu_e$ turn neutrons into protons, and $\bar\nu_e$ protons into neutrons.
We therefore follow neutrinos and antineutrinos of the three flavors separately, with flavor transitions given by probabilities averaged over oscillation phases and with mixing angles in the CP-symmetric thermal potential of the plasma~\cite{Akita:2026xwu}, the same for neutrinos and antineutrinos.
We neglect the refraction from the lepton asymmetry of the emitted particles; for population 3, it is about $0.3\%$ of the vacuum oscillation frequency at the peak of the emitted spectrum at $10^3\,\mathrm{s}$ and smaller later.
Binning the Monte Carlo particles in energy gives the phase-space distributions $f_{\nu_\alpha}(E,t)$ and $f_{\bar\nu_\alpha}(E,t)$ of the flavors $\alpha=e,\mu,\tau$, which set the neutrino rates of the nuclear network (Sec.~\ref{sec:fragments}).
Electrons and positrons heat the plasma and initiate electromagnetic cascades (Sec.~\ref{sec:cascades}).
Muons lose energy in the plasma and decay with relativistic time dilation, and the electron and the two neutrinos of each decay have correlated kinematics.

The energy density still held by the decaying particles, or by the solitons whose mass decreases as they evaporate, enters the Hubble rate together with the neutrinos, the plasma of photons and electrons, and the baryons.
The final baryon-to-photon ratio $\eta_B=n_B/n_\gamma$ is $\eta_B^{\rm CMB}=6.13\times10^{-10}$~\cite{Planck:2018vyg}, and conservation of comoving baryon number gives its earlier values,
\begin{equation}
 \eta_B(t)=\eta_B^{\rm CMB}
 \frac{a_f^3 n_\gamma(t_f)}{a(t)^3n_\gamma(t)} ,
 \label{eq:eta}
\end{equation}
where $t_f$ is any time after the end of photon heating and $a_f=a(t_f)$.
Photons produced after nuclear burning therefore raise $\eta_B$ while nuclei form, relative to standard BBN at the same photon temperature.

\subsection{Electromagnetic cascades}\label{sec:cascades}
Energetic electrons, positrons, and photons interact with the thermal plasma: electrons and positrons through inverse Compton scattering, and photons through pair production on thermal photons, photon-photon scattering, Compton scattering, and pair production on nuclei.
These processes are much faster than the expansion, so the nonthermal photon spectrum at each time is the stationary response to the injection.
Pair production on thermal photons removes cascade photons above $E_c\simeq m_e^2/(22T_\gamma)$; near and below $E_c$, the spectrum depends on the injected energies and differs from the universal power law~\cite{Poulin:2015opa,Hufnagel:2018bjp}.
We compute it by solving the cascade equations with \textsc{Acropolis}~\cite{Depta:2020mhj}, separately for photon and electron injection.
With $S_a(E_0,t)$ the number of primaries $a$ injected per volume, time, and energy, and $G_{\gamma a}(E,E_0)$ the stationary photon spectrum per unit injection rate of primaries $a$ with energy $E_0$, the nonthermal photon number density per unit energy is
\begin{equation}
 F_\gamma(E,t)=\sum_{a=\gamma,e^-,e^+}
 \int dE_0\,G_{\gamma a}(E,E_0;T_\gamma(t))S_a(E_0,t).
 \label{eq:cascade}
\end{equation}
For electrons and positrons, we convolve the response with their injected kinetic-energy spectrum.
Integrating $cF_\gamma$ against each photodisintegration cross section gives the destruction rates.

\subsection{Nuclear network}\label{sec:fragments}
Nuclei made by energetic particles can react in the plasma or be broken up by photons, so the abundances $Y_i=n_i/n_B$ evolve in one network,
\begin{equation}
 \dot Y_i=\mathcal R_i^{\rm nuc}(\boldsymbol Y,T_\gamma,n_B)
 +\mathcal R_i^\nu(\boldsymbol Y,\{f_{\nu_\alpha},f_{\bar\nu_\alpha}\})
 +\mathcal R_i^{\rm frag}(\boldsymbol Y,t)
 +\sum_j\Gamma_{j\to i}^{\gamma}Y_j
 -Y_i\sum_j\Gamma_{i\to j}^{\gamma} .
 \label{eq:network}
\end{equation}
The thermal rates $\mathcal R_i^{\rm nuc}$ use the nuclear reaction rates of Refs.~\cite{Pitrou:2018cgg,Pitrou:2020etk} in their October 2024 tabulation, with finite nucleon mass corrections to weak conversion.
The term $\mathcal R_i^\nu$ contains the conversion of nucleons by the energetic neutrinos and antineutrinos, $\bar\nu_ep\to ne^+$ and $\nu_en\to pe^-$, including muon production above its threshold, and the breakup of $^4$He by neutrinos, $\mathcal R_i^{\rm frag}$ the reactions of the breakup fragments, and $\Gamma^\gamma$ the photodisintegration rates of Sec.~\ref{sec:cascades}.
The network follows $n$, $p$, D, T, $^3$He, $^4$He, $^6$Li, $^7$Li, and $^7$Be.
For the nucleon conversion, we fold each charged-current cross section with the neutrino spectra at each time, with the neutron and proton multiplicities of quasielastic scattering, pion production, and other inelastic channels.
Photodisintegration of the nuclei with mass number three replenishes deuterium, and the final $^3$He/H and $^7$Li/H include the surviving T and $^7$Be.

Lithium-6 forms mainly after energetic neutrinos and fast neutrons break up $^4$He into tritium and $^3$He, which can fuse with ambient helium through $\mathrm T(\alpha,n)^6\mathrm{Li}$ and ${}^3\mathrm{He}(\alpha,p)^6\mathrm{Li}$ before they slow down~\cite{Kawasaki:2017bqm}.
If few nuclear interactions occur while it decelerates, a fragment born at energy $E_0$ produces $^6$Li with probability
\begin{equation}
 P_{A\alpha\to6}(E_0,t)=
 \int_{E_{\rm th}}^{E_0}
 \frac{n_\alpha(t)\,\sigma_{A\alpha\to6}(E)\,v_A(E)}
 {|\dot E_A(E,t)|}\,dE ,
 \label{eq:fragment}
\end{equation}
where $A=\mathrm T$ or ${}^3\mathrm{He}$, $n_\alpha$ is the helium number density, $E_{\rm th}$ is the threshold, and $v_A$ and $\dot E_A$ are the fragment speed and energy-loss rate.
Integrating this probability over the production spectrum of the fragments gives their contribution to $^6$Li.
For the breakup by neutrinos, we fold the shell-model cross sections of Ref.~\cite{Yoshida:2008zb} for the first of its two Hamiltonians for $^4$He, WBP, for each final state with $f_{\nu_\alpha}$ and $f_{\bar\nu_\alpha}$ of all three flavors; they are tabulated up to $E_\nu=180\,\MeV$, and we continue them above in proportion to $E_\nu^2/(E_\nu^2+\Lambda^2)$, with $\Lambda=1\,\mathrm{GeV}$, as in Ref.~\cite{Chang:2024mvg}.
The fragment recoil follows the triton-proton momentum distribution in $^4$He from variational Monte Carlo with the Argonne $v_{18}$ and Urbana~X potentials~\cite{Wiringa:2013ala}, with two prescriptions: the first imposes the kinematically allowed energy on these momenta, and the second also weights the phase space of the final state.

Cascade photons break up $^4$He only once $E_c$ exceeds the $^4$He threshold, below $T_\gamma\simeq0.6\,\keV$, after the lithium solutions stop emitting.
For population 3, the network without the fusion of the helium fragments from neutrino breakup gives $^6$Li/H only about $7\times10^{-15}$ above standard BBN, $1$ to $2\%$ of the $^6$Li from these fragments.

Besides $^6\mathrm{Li}(p,\alpha)^3\mathrm{He}$ and $\mathrm D(\alpha,\gamma)^6\mathrm{Li}$, the network contains
\begin{equation}
\begin{gathered}
 {}^6\mathrm{Li}(n,\alpha)\mathrm T,\qquad
 {}^3\mathrm{He}(\mathrm T,\gamma){}^6\mathrm{Li},\qquad
 \alpha+n+p\longrightarrow{}^6\mathrm{Li}+\gamma,\\
 {}^6\mathrm{Li}(n,\gamma){}^7\mathrm{Li},\qquad
 {}^6\mathrm{Li}(\mathrm D,p){}^7\mathrm{Li},\qquad
 {}^6\mathrm{Li}(\mathrm D,n){}^7\mathrm{Be},
\end{gathered}
\label{eq:li6-reactions}
\end{equation}
with rates and inverse coefficients from \textsc{Primat}~\cite{Pitrou:2018cgg}.
The photon channel $^6\mathrm{Li}(\gamma,\mathrm D){}^4\mathrm{He}$ has a low threshold, $1.47\,\MeV$, and a suppressed electric dipole strength~\cite{Tilley:2002vg}.
We include its measured $2.186\,\MeV$ resonance, which reduces $^6$Li by $0.2$ to $0.3\%$, and omit its nonresonant continuum, which would lower it further.
As the plasma cools, proton burning becomes ineffective, and most of the $^6$Li produced later survives the subsequent neutron and photon exposure.

\subsection{Accuracy of the calculation}\label{sec:numerics}
The allowed populations occupy narrow intervals.
\dh must lie within $2.3\%$ of its measured value and \yp within $0.0026$ of its central value, Eq.~\eqref{eq:S-abundances}, and the cancellation in deuterium holds only within a few percent of the peak lifetime.
The numerical errors must therefore stay well below these widths.

Ref.~\cite{Akita:2026xwu} validates the collision calculation of \nudsmc.
In each calculation, the emitted electrons and neutrinos carry the energy and the charge that the source loses.
To test the injection and redshift of the energetic antineutrinos, which convert protons into neutrons, we compare their energy in the transport with the emitted energy redshifted from the emission times.
From $300\,\mathrm{s}$ on, the two agree within $0.6\%$, and collisions pass at most a few tenths of a percent of this energy to the thermal neutrinos.
Table~\ref{tab:numerics} lists the tests of the remaining numerical choices.
Shifts are differences of the mean abundances between the two settings, and errors are standard errors of these differences.

\begin{table}[t!]
\centering
\footnotesize
\setlength{\tabcolsep}{4pt}
\begin{tabular}{@{}p{0.15\textwidth}p{0.29\textwidth}p{0.2\textwidth}p{0.08\textwidth}p{0.22\textwidth}@{}}
\toprule
Test & Populations ($\epsilon=-1$) & Variation & Simulations & Change of the abundances\\
\midrule
Monte Carlo sampling & $\omega_{\rm peak}=34\,\MeV$, $b=1$, $R_Q=4.4\times10^{-4}$; $\omega_{\rm peak}=45\,\MeV$, $b\approx0.3$, $R_Q=1.4\times10^{-4}$ & minimum number of cells $8\to16$ or $32$; half the target collisions per cell; $2N_\nu$ & $4$ per setting & \dh, \li, \hethree: at most $0.5\%$; \yp: at most $4\times10^{-4}$; \neff: at most $0.005$\\[3pt]
Sampling at the \yp limit & population 3, and the same population at the upper limit of \yp, $N_\nu=2\times10^6$ & as above & $8$ per setting & \yp: below $10^{-4}$; \neff: about $0.001$ or less; \li: at most $0.3\%$\\[3pt]
Discrete charges & $\omega_{\rm peak}=34\,\MeV$, $b\approx0.3$ & $64$ instead of $256$ charges & $10$ each & \dh: $(-0.1\pm0.2)\%$; \li: $(0.4\pm0.5)\%$; \yp: $(1\pm1)\times10^{-4}$; \neff: $-0.002\pm0.002$\\[3pt]
Photodisintegration kernel & population 3; other populations of Table~\ref{tab:soft-points} & $480$ bins instead of the extrapolation & -- & \dh: $+1.9\%$ ($+1.7$ to $+2.1\%$); \li: $+0.2\%$\\
\bottomrule
\end{tabular}
\caption{Shifts of the abundances and \neff of Q-ball populations caused by changing the collision sampling in \nudsmc, the number of discrete charges that represent the charge distribution, and the energy grid of the photodisintegration kernel; the fourth column gives the number of independent simulations per setting. With the refined sampling, population 3 stays below the upper limit of \yp, and the shifts at this limit are consistent with zero within twice their statistical errors.}
\label{tab:numerics}
\end{table}

The second row of Table~\ref{tab:numerics} repeats the sampling test for Q-ball population 3, whose \yp lies closest to the upper limit.
Each discrete charge ends its evaporation with a burst of energetic emission; with too few charges, these bursts appear as small steps in the late destruction of deuterium, so all calculations use $256$ charges.
The photodisintegration kernel is the fraction of a nucleus destroyed per unit energy injected per background photon; it combines Eq.~\eqref{eq:cascade} with the cross sections on $n$ logarithmic energy bins per factor of ten in energy.
From $n=240$ to $960$, each doubling of $n$ roughly halves the change in the destruction rates of D and $^3$He, so we extrapolate to infinitely fine bins from $480$ and $960$ bins; the direct $480$-bin kernel underestimates deuterium destruction.
With either kernel and both recoil prescriptions, \yp and $(^3\mathrm{He}+\mathrm T)/\mathrm H$ of population 3 satisfy their constraints.

Nuclear-rate uncertainties, with that of the CMB baryon density, give standard BBN errors of $1.5\%$ in \dh, $4\%$ in \li, and $1.4\times10^{-4}$ in \yp~\cite{Pitrou:2020etk}.
If they change the abundances with the Q-balls by the same fractions, the error of \dh moves the cancelling lifetime by less than $1\%$, and that of \li is small compared with the width of the stellar band.

\subsubsection{Cross sections for the breakup of helium-4}\label{sec:he4xs}
The nuclear inputs of the fragment calculation (Sec.~\ref{sec:fragments}) make the computed $^6$Li uncertain by a factor of two to three, mostly through the energy spectrum of the recoiling nuclei.
The two prescriptions for this spectrum differ in $^6$Li about twofold and leave the other abundances unchanged, so Tables~\ref{tab:solutions} and \ref{tab:soft-points} give $^6\mathrm{Li}/{}^7\mathrm{Li}$ as the range between them.

The second Hamiltonian of Ref.~\cite{Yoshida:2008zb}, SPSDMK, breaks $^4$He into tritium and helium-3 more often, and its effect grows with the neutrino energy.
For the softer Q-ball populations, it raises $^6$Li by up to about a half and leaves the other abundances nearly unchanged; for the hardest, near $\omega_{\rm peak}=120\,\MeV$, it raises $^7$Li by up to $6\%$, and \hethree can reach its bound.
The particles of Table~\ref{tab:solutions} emit $\nu_e$ and $\bar\nu_e$ of about $130\,\MeV$, and their \hethree stays below its bound with either interaction.
With the lifetime and abundance refitted, the lowest $^7$Li that the particles reach then rises by $2$ to $6\%$ where \neff or \yp sets the limit, below about $265\,\MeV$ for the HNL and $280\,\MeV$ for the gauge bosons; above, \hethree sets the limit, and the minimum rises to $0.14$ to $0.21$ dex above the range of Ref.~\cite{ParticleDataGroup:2026bbn} at $300$ to $320\,\MeV$ (dashed curves in Fig.~\ref{fig:qball-abundances} of the main text).
The lowest $^7$Li of each particle over its masses rises by at most $0.04$ dex, to $2.05$, $2.33$, and $2.36\times10^{-10}$ for the HNL, the $L_e-L_\tau$ boson, and the $B-3L_e$ boson, which thus need at most $0.08$ dex of stellar depletion with either interaction; for the HNL mixing with $\nu_\tau$, \hethree sets the limit already at $350\,\MeV$, and its lowest $^7$Li rises from $2.50$ to $2.87\times10^{-10}$.

Omitting the continuation of the cross sections above $180\,\MeV$ lowers $^6$Li of Q-ball populations 2 and 3 by $13$ to $37\%$, and varying $\Lambda$ between $0.5$ and $2\,\mathrm{GeV}$ changes it by at most $3\%$.
Within this uncertainty, the particle models of Table~\ref{tab:solutions} give $^6\mathrm{Li}/{}^7\mathrm{Li}$ several times the stellar upper limits, and the Q-ball populations of Table~\ref{tab:solutions} give values near or below them, at most about $2\%$; populations with $\omega_{\rm peak}\gtrsim90\,\MeV$ give $9$ to $25\%$.
A calculation of the differential weak response of $^4$He would reduce the uncertainty.

\section{Previous solutions of the lithium problem}\label{sec:single}
In this section, we review previous solutions of the lithium problem and test them against the present measurements, Eqs.~\eqref{eq:S-abundances} and~\eqref{eq:S-lithium}, and other observations.
These solutions lowered $^7$Li with particles of a few MeV~\cite{Pospelov:2010hj,Poulin:2015woa,Ishida:2014wqa,Salvati:2016jng,Berger:2016vxi,Goudelis:2015wpa}, with particles decaying into muon pairs~\cite{Pospelov:2010cw} or hadrons~\cite{Jedamzik:2004er,Kawasaki:2017bqm}, by changing the conditions of nucleosynthesis~\cite{Yamazaki:2014fja,Luo:2018nth,Hou:2017uap,Clara:2020efx,Mori:2019cfo,Makki:2019zem,Talukdar:2024qny}, or with Q-balls that release baryons~\cite{Wang:2004aq,Ichikawa:2004pb}, and they were obtained with the \dh and CMB measurements of their time.
Each subsection states whether newer measurements, effects that we add, or later particle and astrophysical bounds change the verdict.
Mechanisms that change only $^7$Li after nucleosynthesis, such as cosmic strings that disintegrate it~\cite{Koren:2022axd}, leave the other abundances and \neff unchanged and \dh at its standard value.

\subsection{Solutions with decaying particles}\label{sec:earlier-particles}
\paragraph{Neutron injection.}
Neutrons injected near $10^3\,\mathrm{s}$ convert $^7$Be into $^7$Li, which protons destroy, but they also raise \dh through $p(n,\gamma)\mathrm{D}$.
With neutron injection alone, bringing \li below $1.9\times10^{-10}$ requires \dh at least $40\%$ above the upper end of Eq.~\eqref{eq:S-abundances}~\cite{Coc:2014gia}.

Most previous solutions inject the neutrons through hadrons~\cite{Jedamzik:2004er,Kawasaki:2017bqm}.
Ref.~\cite{Jedamzik:2004er} lowered $^7$Li with thermal neutrons, hadronic decays and annihilations, and evaporating Q-balls that emit quarks, all with \dh well above its present measurement.
A TeV particle that decays into $e^+e^-$ also radiates hadrons, in about one decay in a million.
The neutrons from these hadrons lower $^7$Li at $10^2$ to $10^3\,\mathrm{s}$, before the cascades of the $e^+e^-$ can break up deuterium, so \dh rises; Ref.~\cite{Kawasaki:2017bqm} found such solutions for the $10^5\dh=2.53\pm0.04$ measured then.
Ref.~\cite{Bianco:2026ibb} adds the interconversion of protons and neutrons by the injected hadrons and the radiation of $Z$ bosons in the decays; with the present abundances, the particle abundance that these solutions require exceeds its upper bound by more than two orders of magnitude.

Neutrons also come from antineutrinos that convert protons.
Majorons, scalars with the couplings
\begin{equation}
 \mathcal L=-\tfrac12\lambda_{\alpha\beta}\,\phi\,\overline{\nu^c_\alpha}P_L\nu_\beta+\mathrm{h.c.}
 \label{eq:majoron}
\end{equation}
that decay into neutrino pairs, produce such antineutrinos, and from $1\,\MeV$ to $10\,\mathrm{GeV}$, the measured \dh excludes the parameters at which they bring $^7$Li into the stellar band~\cite{Chang:2024mvg}.

\paragraph{Photons below the deuterium threshold.}
Electromagnetic cascades that break up $^7$Be also destroy deuterium or overproduce $^3$He~\cite{Ellis:2005ii}.
Refs.~\cite{Pospelov:2010hj,Poulin:2015woa} proposed photons between the thresholds of $^7$Be and D, $1.59$ and $2.22\,\MeV$, which avoid this, and a source that emits them between about $8\times10^3$ and $2\times10^6\,\mathrm{s}$ satisfies the BBN and CMB constraints at a suitable abundance~\cite{Kawasaki:2020qxm}; \dh then stays at its standard value.
Such photons need particles of a few MeV, and the following paragraphs compare their minimal models with later bounds.

\paragraph{Axion-like particle coupled to photons.}
A pseudoscalar $a$ with
\begin{equation}
 \mathcal L=\frac{g_{a\gamma}}{4}\,a\,F_{\mu\nu}\tilde F^{\mu\nu}
 \label{eq:alp-photon}
\end{equation}
decays into two photons of energy $m_a/2$, with width $\Gamma=g_{a\gamma}^2m_a^3/(64\pi)$.
Ref.~\cite{Depta:2020zbh} showed that such decays lower $^7$Li without destroying too much deuterium only for masses of $3.17$ to $4.44\,\MeV$, whose photons lie between the thresholds.
The couplings that give the required lifetimes conflict at face value with the absence of a gamma-ray signal from SN~1987A~\cite{Jaeckel:2017tud,Muller:2023vjm}, a limit whose robustness is debated~\cite{Depta:2020zbh}; energy deposition in low-energy supernovae restricts the shorter lifetimes~\cite{Caputo:2022mah}, and a global analysis including CMB spectral distortions finds no region that resolves the $^7$Li problem~\cite{Balazs:2022tjl}; see also the nucleosynthesis and CMB bounds of Ref.~\cite{EscuderoAbenza:2025tsi}.
Heavier particles emit photons above the deuterium threshold, where the deuteron cross section soon exceeds that of $^7$Be, so lowering $^7$Li threefold destroys a large fraction of the deuterium~\cite{Kawasaki:2020qxm,Depta:2020zbh}.

\paragraph{Sterile neutrinos.}
A heavy neutral lepton $N$ that mixes with $\nu_\alpha$ through $U_\alpha$ couples to the weak gauge bosons as
\begin{equation}
 \mathcal L=-\frac{g_W}{\sqrt2}\,U_\alpha\,\bar\ell_\alpha\gamma^\mu P_LN\,W^-_\mu-\frac{g_W}{2c_W}\,U_\alpha\,\bar\nu_\alpha\gamma^\mu P_LN\,Z_\mu+\mathrm{h.c.}
 \label{eq:hnl-mixing}
\end{equation}
Ref.~\cite{Ishida:2014wqa} proposed such a lepton of about $14\,\MeV$, whose electrons and positrons produce photons that break up $^7$Be; the destruction of deuterium limits the reduction of $^7$Be to less than a factor of three.
Ref.~\cite{Salvati:2016jng} proposed a lepton near $4.35\,\MeV$, whose rare radiative decays emit photons between the thresholds; its decays into three neutrinos raise \neff to $3.3$ to $3.4$, above Eq.~\eqref{eq:S-lithium}, and its authors regard supernova bounds as disfavoring it.

\paragraph{Scalar mixing with the Higgs boson.}
A scalar that mixes with the Higgs boson couples to fermions in proportion to their masses,
\begin{equation}
 \mathcal L=-\sin\theta\,\phi\sum_f\frac{m_f}{v}\,\bar ff.
 \label{eq:higgs-scalar}
\end{equation}
Ref.~\cite{Berger:2016vxi} found that near $3.5\,\MeV$, where it decays into $e^+e^-$, it lowers $^7$Li; constraints on \neff and on CMB spectral distortions exclude this region~\cite{Fradette:2018hhl}.

\paragraph{Light particles absorbed by nuclei.}
Light particles absorbed by $^7$Be, or freeing neutrons from deuterium near $50\,\keV$, reduce $^7$Li without photodisintegration~\cite{Goudelis:2015wpa}.
The freed neutrons convert $^7$Be and are recaptured into deuterium, which avoids the deuterium excess of an additional neutron source.
The proposed model is an axion-like particle $a$ of a few MeV coupled to $d$ quarks,
\begin{equation}
 \mathcal L=\frac{\partial_\mu a}{f_d}\,\bar d\gamma^\mu\gamma_5d,
 \label{eq:alp-quark}
\end{equation}
with a decay constant $f_d$ of a few TeV, and it needs further interactions to deplete its thermal abundance.
This minimal model, also at other masses and couplings, is in tension with several observations, which we discuss below for masses below and above $5.49\,\MeV$, the threshold of the solar reaction $p(d,{}^3\mathrm{He})a$.

\emph{Masses below $5.49\,\MeV$.}
The Sun emits $a$ in $p(d,{}^3\mathrm{He})a$, and Borexino excludes $m_a=5\,\MeV$ with $f_d=3\,\mathrm{TeV}$~\cite{Borexino:2012guz,BOREXINO:2025dbp}.
With the proposed couplings, the particle escapes the Sun, and Borexino allows only masses just below the threshold, where the slow particles decay before reaching Earth.
Their decays $a\to\gamma\gamma$ then produce photons near $m_a/2\simeq2.7\,\MeV$ that the RHESSI satellite would detect.
We compute this signal with the production rate of Ref.~\cite{Borexino:2012guz} and the M1 capture strength of Ref.~\cite{Anh:2024bmz} at a single solar temperature of $1\,\keV$, including the boosted photon energies and directions and occultation by the Sun, within $0.2\,\mathrm{AU}$ and a $1^\circ$ cone.
For $m_a=5.480$ to $5.485\,\MeV$ and lifetimes of $400$ to $500\,\mathrm{s}$, its peak in $20\,\keV$ bins is three to five times the total detector background, so the criterion of Ref.~\cite{Gustafson:2023hvm} excludes these masses.

Kaon decays constrain the same couplings.
At $m_a=5\,\MeV$ and $f_d=3\,\mathrm{TeV}$, the chiral branching ratio of $K^+\to\pi^+a$~\cite{Bauer:2021wjo,Cornella:2023kjq} with the undetermined weak low-energy constants set to zero exceeds the NA62 limit~\cite{NA62:2025upx} about tenfold; other values of these constants can cancel it.

\emph{Masses above $5.49\,\MeV$.}
These masses escape the solar bounds.
Because mixing with $\pi^0$ sets the width of $a$, the lifetimes of the solution require $f_d$ of about $4$ to $8\,\mathrm{TeV}$ for $m_a$ of $6$ to $7\,\MeV$.
The couplings at low energies then follow from their renormalization-group evolution~\cite{Bauer:2020jbp} and depend on those at high energies.
\begin{itemize}
\item If $a$ couples at high energies only to right-handed $d$ quarks, with all other couplings zero, the running induces an electron coupling that makes it decay into $e^+e^-$ within seconds.
Restoring the lifetime with a larger $f_d$ weakens the absorption so much that the required abundance would heat the photons by several times their energy.
\item Equal couplings to all right-handed down-type quarks, again with no other couplings, induce a flavor-changing interaction whose $K^+\to\pi^+a$ rate exceeds the NA62 limit unless a weak-interaction matrix element is two orders of magnitude above its estimates~\cite{Cornella:2023kjq}.
\item Flavor-dependent charges that are not aligned with the mass basis produce these decays at tree level; avoiding them requires an alignment to about $10^{-9}$.
\end{itemize}
At all masses, depleting the thermal abundance of $a$ into new light particles adds radiation during nucleosynthesis and raises \yp above its upper limit.

\emph{Other mediators.}
A scalar coupled to $d$ quarks frees fewer neutrons than the pseudoscalar, since its coupling to a neutron-proton pair is the same for the bound and free states, and only recoil and the quark-mass dependence of the nuclear force break up the deuteron.
For vectors, the coupling that frees the neutrons also enters their lifetime, production, and laboratory signals in ways that depend on the completion, so a general conclusion would require specifying it.

\paragraph{Muon pairs.}
A particle decaying into two muons with a lifetime of order $10^4\,\mathrm{s}$ raises \dh through the $\bar\nu_e$ from the muons, and cascades from their electrons then remove the excess~\cite{Pospelov:2010cw}.
That analysis lowered $^7$Li by about $60\%$ with \dh bounded only from above, when \neff was still loosely measured.
With the present constraints, and with the dilution of baryons by the energy the particle releases, neutron capture and photodisintegration still cancel in \dh near the proposed lifetime, but $^7$Li reaches the stellar band only where \neff and \yp both exceed Eq.~\eqref{eq:S-abundances}.
Neutrinos carry most of the power of muon decays, and only the $\bar\nu_e$ among them convert protons, so the decays that make enough neutrons raise \neff too much.
Most decays also occur after ${}^7\mathrm{Li}(p,\alpha){}^4\mathrm{He}$ stops, near $3\times10^3\,\mathrm{s}$, when a neutron captured on $^7$Be only turns it into $^7$Li.
Heavier particles emit muons that decay in flight, which does not change this: for masses of $300$ and $500\,\MeV$, \neff reaches its upper limit with \li still $60$ to $80\%$ above the stellar band, and at $1\,\mathrm{GeV}$, \hethree exceeds its bound while \li is still twice the upper end of the band.

\subsection{Solutions that change the conditions of nucleosynthesis}\label{sec:global}
Proposals that change the expansion rate or the baryon-to-photon ratio during nucleosynthesis, or the nuclear reaction rates, alter every abundance at once.
The expansion rate changes with magnetic fields~\cite{Yamazaki:2014fja,Luo:2018nth}, extra radiation~\cite{Makki:2019zem}, or modified gravity~\cite{Talukdar:2024qny}, and the baryon-to-photon ratio with photon cooling~\cite{Yamazaki:2014fja}; the reaction rates change with varying fundamental constants~\cite{Clara:2020efx,Mori:2019cfo} or non-extensive statistics of the plasma~\cite{Hou:2017uap}.
Table~\ref{tab:global} compares the abundances that these solutions report at their best fits with Eqs.~\eqref{eq:S-abundances} and \eqref{eq:S-lithium}; it tests these solutions, not every parameter choice of their mechanisms.
All of them predate the measurements of \yp and \neff in Eqs.~\eqref{eq:S-abundances} and \eqref{eq:S-lithium}, and each leaves at least one bound by more than the difference between its standard BBN and ours.

\begin{table}[t!]
\centering
\footnotesize
\setlength{\tabcolsep}{4pt}
\begin{tabular}{@{}p{0.08\textwidth}p{0.27\textwidth}p{0.45\textwidth}p{0.12\textwidth}@{}}
\toprule
Ref. & Modification & Reported at the solution & Conflicting observable\\
\midrule
\cite{Yamazaki:2014fja} & photon cooling, with decays and a magnetic field & $\eta_B$ during nucleosynthesis three quarters of its CMB value, so photons lose a third of their energy before recombination and $\neff\approx4.5$ & \neff\\
\cite{Luo:2018nth} & inhomogeneous magnetic field & $\yp=0.2568$, $10^5\dh=3.21$ & \yp, \dh\\
\cite{Hou:2017uap} & non-extensive statistics & $10^5\dh=3.14$ to $3.25$ & \dh\\
\cite{Clara:2020efx} & varying fine-structure constant, three coupling models & unification: $\yp=0.254\pm0.002$, $10^5\dh=2.68\pm0.07$; dilaton and clocks: $10^{10}\li=3.8$ and $4.1$ & \yp; $^7$Li\\
\cite{Mori:2019cfo} & varying quark mass & $^7$Li enters the stellar band only where $10^5\dh\approx2.75$ & \dh\\
\cite{Makki:2019zem} & neutrino properties and dark matter & $\neff=4.3$ to $4.9$, $\yp=0.231$ to $0.240$, $10^5\dh=3.5$ to $3.7$ & \neff, \yp, \dh\\
\cite{Talukdar:2024qny} & $f(R)$ gravity & $\yp=0.250$ to $0.257$, $10^5\dh=8.5$ to $18$ & \yp, \dh\\
\bottomrule
\end{tabular}
\caption{Lithium solutions from the literature that change the conditions of nucleosynthesis (Sec.~\ref{sec:global}), with the abundances that these references report at their best fits; for Ref.~\cite{Mori:2019cfo}, \dh is read from its figure at the quark-mass shift where $^7$Li enters the stellar band of Eq.~\eqref{eq:S-lithium}. The last column names the observables whose values lie outside Eqs.~\eqref{eq:S-abundances} and \eqref{eq:S-lithium}.}
\label{tab:global}
\end{table}

\subsection{Solutions with Q-balls that release baryons}\label{sec:other-sources}
Two earlier proposals lowered $^7$Li with Q-balls that release baryons, either as nucleons during nucleosynthesis~\cite{Wang:2004aq} or after it, diluting the primordial lithium~\cite{Ichikawa:2004pb}.

Q-balls evaporating into nucleons near $10^3\,\mathrm{s}$ keep $\eta_B$ below its CMB value while $^7$Be forms, so that less $^7$Be is made~\cite{Wang:2004aq}.
That study estimated the reduction of $^7$Be from the evolution of $\eta_B$.
The released nucleons also include neutrons, and by our estimate, with $p/n=7$, these Q-balls inject about a thousand times as many neutrons per final baryon as population 3, whose neutrons alone raise \dh by about $60\%$.
As for pure neutron injection~\cite{Coc:2014gia}, \dh then ends far above Eq.~\eqref{eq:S-abundances}.

Baryons released by Q-balls after nucleosynthesis dilute D and $^7$Li relative to hydrogen by the same factor, and a lepton asymmetry $\xi_e\simeq-0.4$ restores helium~\cite{Ichikawa:2004pb}.
That proposal was compared with the higher and less precise \dh measured at the time.
The dilution cannot lower $^7$Li by the required factor without raising \dh too much: while nuclei form, $\eta_B$ is below its final value, so \dh rises as $^7$Li falls, and the neutrons added by the lepton asymmetry raise \dh further.
We compute this with our network for $0$ to $50\%$ of the final baryons released after nucleosynthesis and neutrino degeneracy parameters from $-0.8$ to $0.1$ in $\nu_e$ alone, or from $-0.6$ to $0.1$ equal in all flavors; both optima below lie inside this range.
Where $^7$Li reaches the stellar band and \yp stays within Eq.~\eqref{eq:S-abundances}, \dh exceeds its upper limit by $21\%$; with \dh below this limit, \li falls only $18\%$ below its standard value and stays $1.9$ times the upper end of Eq.~\eqref{eq:S-lithium}.

\section{Decay spectra and particle models for lithium-7 and deuterium}\label{sec:particles}
In this section, we determine which decay spectra bring $^7$Li and \dh to their observed values within Eqs.~\eqref{eq:S-abundances} and~\eqref{eq:S-lithium}, and which particle models produce them.

Decay spectra that work combine energetic antineutrinos with enough electromagnetic emission to remove the extra deuterium; their flavor composition and branching fractions decide which interactions produce them.

\subsection{Decay spectra that bring lithium-7 and deuterium to their observed values}\label{sec:single-hard}
To find which emission compositions work, and to compare one decaying species with the Q-ball populations (Sec.~\ref{sec:emission}), we compute species $X$ that decay at rest into electrons and neutrinos, each with energy $E$.
One third of the emitted particles are $\nu_\mu$ or $\bar\nu_\mu$, a fraction $b$ of the others are electrons and positrons, and the rest are $\nu_e$ and $\bar\nu_e$,
\begin{equation}
 N_{\nu_\mu}:N_{e^\pm}:N_{\nu_e}=1:2b:2(1-b),
 \label{eq:S-composition}
\end{equation}
where $N_{\nu_\mu}$ counts $\nu_\mu$ and $\bar\nu_\mu$, $N_{e^\pm}$ electrons and positrons, and $N_{\nu_e}$ counts $\nu_e$ and $\bar\nu_e$.
This is the composition of the electron populations of Q-balls (Sec.~\ref{sec:source}), whose $\nu_\mu$ third follows from the condensate; a decaying particle need not emit it, and Sec.~\ref{sec:two-particles} computes the particle models with their own decay products.
$X$ emits either the charge-conjugate particles, $\epsilon=-1$, whose unpaired third is $\bar\nu_\mu$, or both kinds equally, $\epsilon=0$.
The electrons and positrons receive the fraction $2b/3$ of the energy, and the neutrinos the rest, which raises \neff.
Already at $b=0.6$, \neff, or \hethree at the highest energies, reaches its bound before $^7$Li enters the stellar band, and smaller $b$ put still more energy into neutrinos.
We therefore compute $b\gtrsim0.3$ for $\epsilon=-1$, and $b=0.75$ for $\epsilon=0$, at energies $E$ up to about $230\,\MeV$.

\begin{figure}[t!]
\centering
\includegraphics[width=0.6\textwidth]{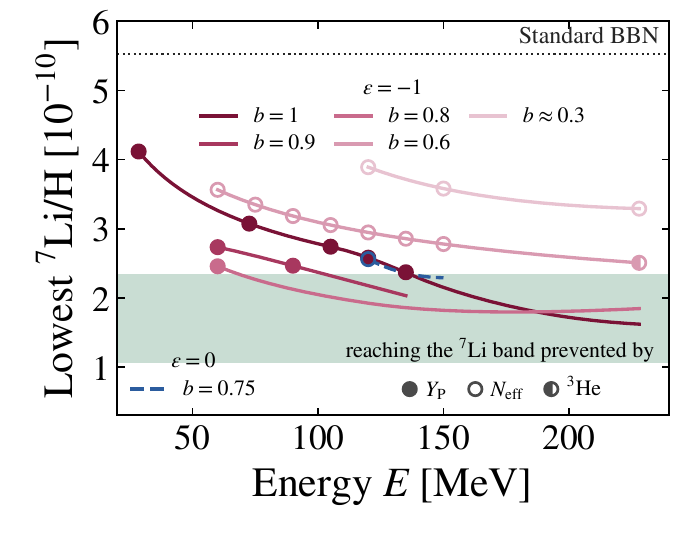}
\caption{Lowest $^7\mathrm{Li/H}$ for one decaying species $X$ versus the energy $E$ of each emitted electron or neutrino, at the lifetime at which $10^5\dh=2.508$, reached when the first bound of Eqs.~\eqref{eq:S-abundances} and \eqref{eq:S-lithium} fails as the abundance of $X$ grows.
$X$ decays with a single lifetime into the electrons and neutrinos of an electron population with charged fraction $b$, Eq.~\eqref{eq:S-composition}, with an excess of antineutrinos, $\epsilon=-1$ (red, darker for larger $b$), or symmetrically, $\epsilon=0$ (blue, dashed); the curves are smooth fits in $\ln E$ that stay within $0.14\times10^{-10}$ of every computed value.
Where $^7$Li cannot reach the stellar band (green), markers show the bound that prevents it, filled for \yp, open for \neff, and half-filled for \hethree; curves inside the band are lithium solutions.
For $\epsilon=0$, $b=1$ at all energies and $b=0.75$ above $150\,\MeV$ exceed the \yp and \hethree bounds already at the smallest computed abundance and have no points.}
\label{fig:single-map}
\end{figure}

For each composition and energy, we fix the lifetime of $X$ at which $10^5\dh=2.508$ and raise its abundance until a bound of Eqs.~\eqref{eq:S-abundances} and \eqref{eq:S-lithium} fails.
Fig.~\ref{fig:single-map} shows the lowest $^7\mathrm{Li/H}$ reached before this happens; $X$ is a lithium solution where this value lies in the stellar band.

We take $X$ nonrelativistic from the freeze-out of neutrons and protons on.
A species still relativistic then would add radiation.
With an energy fraction $f_{\rm rel}$ at that time, it would raise \neff during nucleosynthesis by about $7f_{\rm rel}$ and \yp by about $0.1f_{\rm rel}$~\cite{Pitrou:2018cgg}.
For $f_{\rm rel}$ equal to the energy fractions $f$ of Table~\ref{tab:solutions}, this rise of \yp, $1.4$ to $5.5\times10^{-3}$, would match or exceed the $1.3\times10^{-3}$ between standard BBN and the upper limit of Eq.~\eqref{eq:S-abundances}, before any proton is converted.

Three trends follow (Figs.~\ref{fig:single-map} and~\ref{fig:family}).
\begin{itemize}
\item Over the computed energies, more energetic emission lowers $^7$Li further, since the neutrons made per unit of injected energy grow with the energy of the antineutrinos.
\item The charged fraction decides which bound fails first.
For $b=0.95$ and $1$, the $e^\pm$ carry most of the energy, and \yp reaches its upper limit first unless $E\gtrsim150\,\MeV$.
For $b=0.8$, $^7$Li enters the band from about $75\,\MeV$ up to at least $230\,\MeV$, and for $b=0.9$ from about $100\,\MeV$.
For $b\leq0.6$, \neff reaches its upper limit first, or \hethree at the highest computed energies.
\item With symmetric emission, neutrinos and antineutrinos come in equal numbers, and the $\nu_e$ partly undo the conversion of protons by the $\bar\nu_e$, so solutions exist only near $140$ to $150\,\MeV$ for $b=0.75$, between the \neff limit at lower energies and the \hethree bound at higher ones.
\end{itemize}

Fig.~\ref{fig:family} places these solutions in the plane of $^6\mathrm{Li}/{}^7\mathrm{Li}$ and \neff, together with the Q-ball lithium solutions; they need energy fractions $f$, Eq.~\eqref{eq:f}, of $3$ to $6\%$, at which \yp lies between $0.2477$ and $0.2483$ and \hethree between $0.86$ and $1.39\times10^{-5}$.

Every solution makes $^6$Li, since its antineutrinos of $75\,\MeV$ or more break up helium into energetic fragments (Fig.~\ref{fig:family}).
Where $^7$Li enters the band, $^6\mathrm{Li}/{}^7\mathrm{Li}$ grows with $E$ from below $1\%$ to over $20\%$.
Only the least energetic solutions reach the range of the stellar upper limits~\cite{Wang:2021urb}.
With $b=0.8$ at $75\,\MeV$, $^6\mathrm{Li}/{}^7\mathrm{Li}$ stays within this range in both recoil prescriptions; at $90\,\MeV$ with $b=0.8$ and $105\,\MeV$ with $b=0.9$, only in the prescription that also weights the phase space of the final state.
All three have $\epsilon=-1$, an excess of antineutrinos over neutrinos, which a boson decaying into $e^+e^-$ and neutrino pairs cannot emit; Sec.~\ref{sec:two-particles} gives a fermion that emits it.

\subsection{Particle models that bring lithium-7 and deuterium to their observed values}\label{sec:two-particles}
The following models produce decay spectra of Sec.~\ref{sec:single-hard} that bring \dh and $^7$Li to their observed values while \yp, $^3$He, and \neff stay within their bounds (Table~\ref{tab:realizations}).
They decay at rest, and their masses and couplings fix the spectra of the emitted electrons, neutrinos, and photons; two-body decays give lines.
Lifetimes near $10^4\,\mathrm{s}$ require couplings so weak that these particles stay far from equilibrium at temperatures near their masses, where inverse decays make only about $2\times10^{-4}$ of the required abundance.
As for the Q-balls, we therefore treat the abundance as an independent parameter.
Scattering through the dipole coupling of the HNL grows with temperature, so its yield depends on the earlier thermal history.
The asymmetry of the fermion $\chi$ below must also come from an earlier process.
Section~\ref{sec:supernovae} describes how supernovae and neutron-star mergers produce them and which limits apply to the resulting signals, among them the photons recorded after SN~1987A by the Solar Maximum Mission (SMM) and the Pioneer Venus Orbiter (PVO); the paragraphs below give each signal relative to its limit for the cold and the hot core.

\begin{figure}[t!]
\centering
\includegraphics[width=0.6\textwidth]{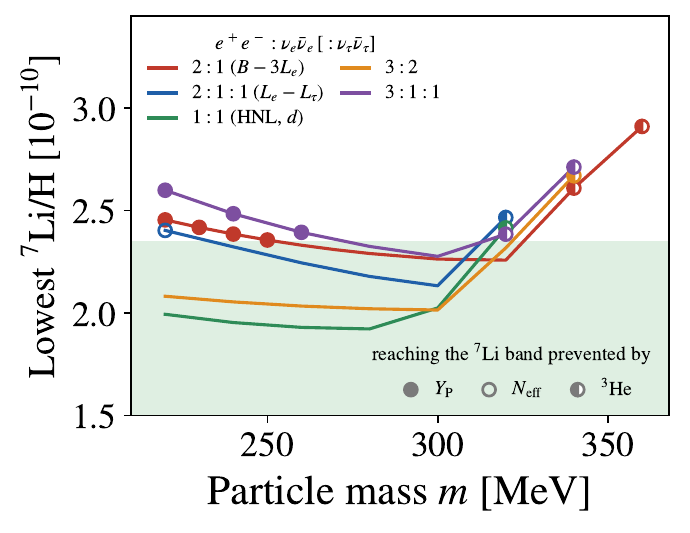}
\caption{Lowest $^7$Li/H versus the mass $m$ of a particle that decays at rest into $e^+e^-$ and neutrino pairs, with \dh at its measured value and the other abundances within Eq.~\eqref{eq:S-abundances} and $\neff\leq3.161$.
The legend gives the ratio of $e^+e^-$ to $\nu_e\bar\nu_e$ (to $\nu_\tau\bar\nu_\tau$) pairs and the models with this ratio; for the HNL, the photon counts as $e^+e^-$, since it produces an $e^+e^-$ pair on background photons almost at once and starts the same cascade.
The curves are smooth fits of the simulations for each ratio.
Where $^7$Li cannot reach the stellar band of Eq.~\eqref{eq:S-lithium} (green), markers show the bound that prevents it, filled for \yp, open for \neff, and half-filled for \hethree; inside the band, the particles are lithium solutions.}
\label{fig:region-particles}
\end{figure}

\paragraph{Gauge boson of $L_e-L_\tau$.}
The boson of Eq.~\eqref{eq:couplings} with the charges of $L_e-L_\tau$~\cite{He:1990pn,He:1991qd,Foot:1990mn} decays into $e^+e^-$, $\nu_e\bar\nu_e$, and $\nu_\tau\bar\nu_\tau$ in the ratio $2:1:1$, each particle with energy $m/2$, with width $\Gamma=g^2m/(6\pi)$.
Kinetic mixing with the photon, induced by electron and tau loops, adds $\mu^+\mu^-$ and $\pi^+\pi^-$ in at most $10^{-5}$ of the decays.
With these decays, $^7$Li enters the stellar band from about $240$ to $310\,\MeV$ before \neff reaches its limit; at higher masses, \hethree exceeds its bound first (Fig.~\ref{fig:region-particles}).
At $260\,\MeV$, the stellar upper limits on $^6\mathrm{Li}/{}^7\mathrm{Li}$~\cite{Wang:2021urb} require stars to deplete $^6$Li $4$ to $24$ times more than $^7$Li; models of metal-poor stars reach such factors for some stars~\cite{Asplund:2005yt}, but not for two of the three stars that set the limits.

Supernovae produce this boson through its couplings to electrons and neutrinos.
After SN~1987A, at $260$ to $300\,\MeV$, its photons in the SMM band lie $3$ to $6$ times below the limit for a hot core and far below it for a cold one, and its soft photons $2.4$ to $3$ times below the single-channel PVO limit.
Its other signals, from positrons, from the energy deposited inside the progenitor, and from GW170817, stay at least $25$ times below their limits, and laboratory searches reach it only for $g\gtrsim10^{-8}$~\cite{Bauer:2018onh}.

\paragraph{Gauge boson of $B-3L_e$.}
A right-handed neutrino of charge $-3$ cancels the cubic and gravitational anomalies that $e$ and $\nu_e$ leave in $B-3L_e$; those with the gauge groups of the Standard Model cancel, as for $B-L$~\cite{Ma:1997nq,Chang:2000xy}.
We take it heavier than $m/2$, so the boson of Eq.~\eqref{eq:couplings} decays into $e^+e^-$ and $\nu_e\bar\nu_e$ in the ratio $2:1$, with width $\Gamma=9g^2m/(8\pi)$; its coupling to quarks adds $\pi^0\gamma$ in about $10^{-6}$ of the decays.
Kinetic mixing with the photon depends on the scale at which it vanishes; if this scale lies far above $m$, the $e^+e^-$ fraction rises from $2/3$ to at most $0.70$, and $\mu^+\mu^-$ appear in up to $0.5\%$ of the decays above the muon threshold.

With the $2:1$ decays, the boson brings $^7$Li to the upper edge of the stellar band just where \yp reaches its upper limit, from $260$ to $320\,\MeV$ (Fig.~\ref{fig:region-particles}); at $260$ and $270\,\MeV$, four independent simulations give lithium solutions in narrow intervals of the energy fraction (the first in Table~\ref{tab:solutions}), and from $340\,\MeV$ on, \hethree exceeds its bound first.

Its larger coupling to electrons makes supernovae produce it more often than the $L_e-L_\tau$ boson.
After SN~1987A, its photons in the SMM band lie only $2.4$ to $2.9$ times below the limit for a hot core, though far below it for a cold one, and its soft photons $1.8$ times below the single-channel PVO limit.
In the $1$ to $3\,\MeV$ channel, for which we estimate the effective area, they would reach it.
Its other signals stay at least $18$ times below their limits, and laboratory searches reach only far larger couplings.

\paragraph{Gauge bosons of $B-xL_e-(3-x)L_\tau$.}
With three right-handed neutrinos heavier than $m/2$ and charges $-x$ and $-(3-x)$ for the electron and tau leptons, any $x$ cancels the anomalies; $x=3$ gives $B-3L_e$, and $x=3/2$ the decays of the $L_e-L_\tau$ boson.
Between them, the fraction of decays into $\nu_\tau\bar\nu_\tau$ falls from a quarter to zero, so that \yp ends the reduction of $^7$Li before \neff reaches its limit.

For $x=1.6$ to $1.8$ near $265\,\MeV$, $^7$Li reaches $(2.1$ to $2.2)\times10^{-10}$, or $(2.2$ to $2.3)\times10^{-10}$ with the SPSDMK Hamiltonian for $^4$He breakup, with \yp near its upper limit, and the photons in the SMM band stay at least $2.4$ times below the limit.

\paragraph{Heavy neutral lepton with a dipole coupling to $\nu_e$.}
The HNL with the dipole coupling $d_e$ of Eq.~\eqref{eq:couplings}~\cite{Magill:2018jla} decays into $\nu_e\gamma$ and $\bar\nu_e\gamma$, each product with energy $m/2$, with width $\Gamma=d_e^2m^3/(4\pi)$ for a Dirac fermion and twice this for a Majorana fermion; three-body decays through a virtual photon make less than $1\%$ of them.
Its photon produces an $e^+e^-$ pair on background photons almost at once, and the resulting photodisintegration of D and $^7$Be differs by only about $1\%$ from that of injected $e^\pm$ of the same energy (\textsc{Acropolis}~\cite{Depta:2020mhj}).

The HNL brings $^7$Li into the stellar band from $220\,\MeV$, the lowest computed mass, to about $310\,\MeV$, above which \hethree exceeds its bound first (Fig.~\ref{fig:region-particles}).

The dipole also induces mass mixing of $N$ with $\nu_e$, whose decays must stay below $1\%$ of those through the dipole, $U_e^2\lesssim5\times10^{-16}$; this requires a cutoff of the dipole operator below about $5\,\mathrm{TeV}$, or charged states in its completion below about $110\,\mathrm{GeV}$, or a tuning of about $1\%$ for charged states near $1\,\mathrm{TeV}$~\cite{Magill:2018jla}.

After SN~1987A, its photons in the SMM band lie far below the limit for the cold core; for the hot core, they lie an order of magnitude below it at $300\,\MeV$ but close to it at $265$ to $275\,\MeV$.
Most of its photons arrive above $100\,\MeV$, outside this band. With a rough estimate of the response of the detector to them, the cold-core signal stays below the limit at $300\,\MeV$ and, except for the colder core with the larger electron-neutrino chemical potential, at $265$ to $280\,\MeV$, whereas the hot-core signal exceeds it.

Above about $260\,\MeV$, the HNL lies outside the region that the dedicated analysis of these data excludes~\cite{Brdar:2023tmi}.
Its other signals stay at least $30$ times below their limits, except its photons from GW170817 for a remnant at a uniform $50\,\MeV$, which lie $4$ times below the Fermi-LAT limit, and laboratory searches reach only far larger couplings~\cite{Magill:2018jla}.

\paragraph{Scalar coupled to electrons and neutrinos.}
A real scalar with
\begin{equation}
 \mathcal L=-g_e\,\phi\,\bar ee-\left(\tfrac12\lambda_{ee}\,\phi\,\overline{\nu^c_e}P_L\nu_e+\tfrac12\lambda_{\tau\tau}\,\phi\,\overline{\nu^c_\tau}P_L\nu_\tau+\mathrm{h.c.}\right)
 \label{eq:scalar-enu}
\end{equation}
and couplings that give $e^+e^-$, $\nu_e\nu_e$ or $\bar\nu_e\bar\nu_e$, and $\nu_\tau\nu_\tau$ or $\bar\nu_\tau\bar\nu_\tau$ in the ratio $2:1:1$ reproduces the decays of the $L_e-L_\tau$ boson and its lithium solutions.
Its couplings to neutrinos violate lepton number, and a coupling to muons must stay below $(0.3$ to $0.4)\,g_e$.

Its coupling to $\nu_e\nu_e$ makes supernovae produce it far more often than the vector boson.
For electron-neutrino chemical potentials of $150\,\MeV$ in the core, its photons in the SMM band lie between five times below and three times above the limit, depending on the core temperature; for $100\,\MeV$, they lie two to five times below it in a hot core.

\paragraph{Asymmetric fermion decaying through a scalar.}
A Dirac fermion $\chi$ that carries the lepton number of $\nu_\mu$ decays through the scalar of Eq.~\eqref{eq:scalar-enu}, with $\lambda_{\tau\tau}=0$, via
\begin{equation}
 \mathcal L=-y\,\phi\,\bar\nu_\mu P_R\chi+\mathrm{h.c.}
 \label{eq:chi}
\end{equation}
It decays as $\bar\chi\to\bar\nu_\mu\phi$, with width $\Gamma_\chi=y^2m_\chi(1-m_\phi^2/m_\chi^2)^2/(32\pi)$; with $m_\chi=3E$ and $m_\phi=\sqrt3E$, the antineutrino has energy $E$, and the decay products of $\phi$ are distributed uniformly between $E/2$ and $3E/2$.
An asymmetric population of $\bar\chi$ thus emits the particles of an electron population with $\epsilon=-1$ and $b$ equal to the $e^+e^-$ branching of $\phi$; the spread of the energies of the decay products of $\phi$ changes the cascade response by at most $0.3\%$ and raises the $E^2$-weighted rates of the neutrino pairs by $8\%$, so $\chi$ approximates the computed species of Fig.~\ref{fig:family}, closely for $b=1$, where $\phi$ emits no neutrino pairs.

Its lithium solutions need a relic $\bar\chi$ asymmetry about $5\times10^4$ times that of baryons, and a coupling to muons below $(0.3$ to $0.5)\,g_e$ where $\phi\to\mu^+\mu^-$ is open.
Laboratory experiments would produce $\chi$ at rates suppressed by $y^2\sim10^{-26}$, and with $g_e$ between about $10^{-5}$ and $10^{-4}$, $\phi$ decays before arriving at the detectors of beam dumps and escapes the search for dark photons at BaBar~\cite{BaBar:2014zli}, recast for a scalar.

\begin{table*}[t!]
\centering
\footnotesize
\renewcommand{\arraystretch}{1.15}
\begin{tabular}{@{}p{0.14\textwidth}p{0.2\textwidth}p{0.15\textwidth}p{0.19\textwidth}p{0.09\textwidth}p{0.17\textwidth}@{}}
\toprule
\raggedright Particle & Decays & $m$ [MeV] & Coupling & $10^2\,{}^6\mathrm{Li}/{}^7\mathrm{Li}$ & Further requirements \tabularnewline
\midrule
\raggedright $Z'$ of $L_e-L_\tau$ & $e^+e^-$, $\nu_e\bar\nu_e$, $\nu_\tau\bar\nu_\tau$ as $2:1:1$ & $240$ to $310$ & $g\approx7\times10^{-14}$ & $6.1$ to $15$ & None \tabularnewline
\raggedright $Z'$ of $B-3L_e$ & $e^+e^-$, $\nu_e\bar\nu_e$ as $2:1$ & $260$ to $320$ & $g\approx3\times10^{-14}$ & $5$ to $13$ & Small kinetic mixing at $m$ \tabularnewline
\raggedright HNL, dipole $d_e$ & $\nu_e\gamma$, $\bar\nu_e\gamma$ & $220$ to $310$ & $d_e=(1.2$ to $2.1)\times10^{-13}\,\mathrm{GeV}^{-1}$ (Majorana) & $7.1$ to $17$ & Mass mixing $U_e^2\lesssim5\times10^{-16}$ \tabularnewline
\raggedright $\phi$; $g_e$, $\lambda_{ee}$, $\lambda_{\tau\tau}$ & As the $Z'$ of $L_e-L_\tau$ & $240$ to $310$ & $g_e\approx(4$ to $5)\times10^{-14}$ & As the $Z'$ & $g_\mu<(0.3$ to $0.4)\,g_e$ \tabularnewline
\raggedright $\bar\chi$ and $\phi$; $y$, $g_e$ & $\bar\chi\to\bar\nu_\mu\phi$, $\phi\to e^+e^-$ & $m_\chi=450$ to $684$, $m_\phi=260$ to $395$ & $y=(1.6$ to $2.0)\times10^{-13}$ & $5$ to $19$ & $\bar\chi$ asymmetry ${\approx}\,5\times10^4$ times that of baryons \tabularnewline
\raggedright $\bar\chi$ and $\phi$; $y$, $g_e$, $\lambda_{ee}$ & As above, with $10$ to $20\%$ of $\phi\to\nu_e\nu_e,\bar\nu_e\bar\nu_e$ & $m_\chi=270$ to $315$, $m_\phi=156$ to $182$ & $y=(2.4$ to $2.6)\times10^{-13}$ & $0.7$ to $5$ & As above; $g_\mu\lesssim(0.3$ to $0.5)\,g_e$ \tabularnewline
\bottomrule
\end{tabular}
\caption{Decays, masses $m$, couplings, $^6\mathrm{Li}/{}^7\mathrm{Li}$, and further conditions of the particle models of Sec.~\ref{sec:two-particles} that bring \dh and $^7$Li to their observed values while \yp, $^3$He, and \neff stay within their bounds; the couplings give lifetimes of $(8$ to $9)\times10^3\,\mathrm{s}$.
The masses span the lithium solutions of Fig.~\ref{fig:region-particles}; for the HNL, they reach $220\,\MeV$, the lowest computed mass.
$^6\mathrm{Li}/{}^7\mathrm{Li}$ is given at the masses of Table~\ref{tab:solutions} and spans the two recoil prescriptions of Sec.~\ref{sec:fragments}. Table~\ref{tab:solutions} of the main text gives the other abundances for the first three.}
\label{tab:realizations}
\end{table*}

\subsection{Particle models that fail}\label{sec:failing-particles}
Other particles fail because their decays lack energetic antineutrinos or electromagnetic energy, or include muons or charged pions.
Electrons and positrons alone only destroy deuterium, and neutrinos alone only raise it.
Muon pairs give most of their energy to neutrinos and raise \neff before $^7$Li is low enough, and charged pions from the early decays convert protons and raise \yp.
Table~\ref{tab:failing} lists the models that we computed and the leading bound of each; Sec.~\ref{sec:single} discusses the earlier proposals.

\paragraph{Dark photon.}
A dark photon couples to the electromagnetic current $J^\mu_{\rm em}$ through kinetic mixing,
\begin{equation}
 \mathcal L=\epsilon e\,A'_\mu J^\mu_{\rm em}.
 \label{eq:dark-photon}
\end{equation}
Below the muon threshold, it decays only into $e^+e^-$, whose cascades break up $^7$Be only while also destroying deuterium or, through $^4$He, overproducing $^3$He~\cite{Ellis:2005ii}; heavier dark photons add muon and pion pairs.

\paragraph{Gauge bosons of $L_\mu-L_\tau$, $L_e-L_\mu$, and $B-L$.}
These bosons couple as in Eq.~\eqref{eq:couplings}, with $Q=1$ for $\mu$ and $\nu_\mu$ and $Q=-1$ for $\tau$ and $\nu_\tau$ ($L_\mu-L_\tau$), $Q=1$ for $e$ and $\nu_e$ and $Q=-1$ for $\mu$ and $\nu_\mu$ ($L_e-L_\mu$), or $Q=1/3$ for quarks and $Q=-1$ for leptons ($B-L$, with right-handed neutrinos heavier than $m/2$).
The $L_\mu-L_\tau$ boson emits almost no electromagnetic energy below the muon threshold, so the cascades cannot remove the extra deuterium, and muon pairs above it.
The $L_e-L_\mu$ and $B-L$ bosons emit $e^+e^-$, but also $\nu_\mu$ and $\nu_\tau$ pairs, which convert no protons and raise \neff before $^7$Li reaches the stellar band.

\paragraph{Heavy neutral leptons.}
Through mixing with $\nu_e$ or $\nu_\mu$, Eq.~\eqref{eq:hnl-mixing}, a heavy neutral lepton decays mostly into a charged lepton and a charged pion, and these pions raise \yp; through mixing with $\nu_\tau$, it decays mostly into $\nu\pi^0$, and \yp or \neff reaches its limit before $^7$Li enters the band, at $\li\approx2.5\times10^{-10}$ for $350\,\MeV$ (Fig.~\ref{fig:qball-abundances} of the main text); for a cold supernova core, SN~1987A allows it above about $300\,\MeV$~\cite{Chauhan:2025sterileSN}.
With $\nu_\tau$ in place of $\nu_e$ in the dipole coupling of Eq.~\eqref{eq:couplings}, its $\nu_\tau$ and $\bar\nu_\tau$ convert protons only after oscillating into $\nu_e$ and $\bar\nu_e$, and \neff exceeds its upper limit first.

\paragraph{Higgs-like scalar and majorons.}
Below the muon threshold, the scalar of Eq.~\eqref{eq:higgs-scalar} decays into $e^+e^-$ and $\gamma\gamma$, whose cascades alone cannot raise \dh; above it, the scalar decays mostly into muon pairs, which fail as in Sec.~\ref{sec:earlier-particles}.
Majorons with only the couplings of Eq.~\eqref{eq:majoron} fail at all masses (Sec.~\ref{sec:earlier-particles}).
In seesaw models, a coupling to electrons arises at one loop together with a coupling to nucleons more than a thousand times larger, and by our estimate, the signals of SN~1987A then exceed the SMM and $511\,\keV$ limits where the majoron also decays into $e^+e^-$.

\begin{table*}[t!]
\centering
\footnotesize
\setlength{\tabcolsep}{5pt}
\begin{tabular}{@{}llllc@{}}
\toprule
Particle & Decays & $m$ [MeV] & Leading bound & $10^{10}\,{}^7\mathrm{Li/H}$\\
\midrule
$Z'$ of $L_\mu-L_\tau$ & $\nu_\mu\bar\nu_\mu$, $\nu_\tau\bar\nu_\tau$; $\mu^+\mu^-$ above $211\,\MeV$ & below $211$ & \dh stays above Eq.~\eqref{eq:S-abundances} & --\\
 & & $230$ to $330$ & \neff & $\geq3.4$\\
$Z'$ of $L_e-L_\mu$ & $e^+e^-$, $\nu_e\bar\nu_e$, $\nu_\mu\bar\nu_\mu$ as $2:1:1$; $\mu^+\mu^-$ & $230$ to $330$ & \neff; \hethree near $330$ & $\geq2.8$\\
$Z'$ of $B-L$ & $e^+e^-$ and each $\nu\bar\nu$ as $2:1$; $\mu^+\mu^-$ & $150$ to $330$ & \neff & $\geq3.0$\\
HNL, mixing with $\nu_e$ or $\nu_\mu$ & $e^\mp\pi^\pm$, $\mu^\mp\pi^\pm$ & up to $690$ & \yp & --\\
HNL, mixing with $\nu_\tau$ & $\nu\pi^0$ & $300$ to $350$ & \yp, \neff & $2.5$\\
HNL, dipole coupling to $\nu_\tau$ & $\nu_\tau\gamma$, $\bar\nu_\tau\gamma$ & $300$ to $400$ & \neff & --\\
Higgs-like scalar & $e^+e^-$, $\gamma\gamma$; $\mu^+\mu^-$ above $211\,\MeV$ & above $211$ & \neff & $\geq3.5$\\
\bottomrule
\end{tabular}
\caption{Particle models that fail, with their decays, the computed masses, the leading bound, which fails first as the abundance grows at the lifetime at which $10^5\dh=2.508$, and the lowest $^7\mathrm{Li/H}$ reached before it fails.}
\label{tab:failing}
\end{table*}

\section{Supernova and merger signals of the particles}\label{sec:supernovae}
The couplings that give the particles of Sec.~\ref{sec:two-particles} lifetimes near $10^4\,\mathrm{s}$ also produce them in the hot cores of supernovae and in the remnants of neutron-star mergers.
Most of them decay far outside the star, and observations limit the photons, positrons, and energy that their decays release.
In this section, we compute this production and the resulting signals and compare them with these limits.

\paragraph{Production in a supernova core.}
In a proto-neutron star, at temperatures of $30$ to $60\,\MeV$ and near nuclear density, the particles form mainly by coalescence.
Pairs of $e^+e^-$ and $\nu\bar\nu$ make the vector bosons, and neutrinos and photons make the HNL.
The rate of coalescence follows from the decay width and the occupations of the colliding particles, so the lifetime that the lithium solution requires fixes it.
For masses of a few hundred MeV, the factor $e^{-m/T}$ suppresses it.
Compton-like production, pion absorption, and nucleon bremsstrahlung add less than $1\%$ for the $L_e-L_\tau$ boson and $2$ to $4\%$ for the $B-3L_e$ boson.

The core holds more electron neutrinos than antineutrinos.
This excess raises the production of the HNL from $\nu_e\gamma$ and of the scalar of Eq.~\eqref{eq:scalar-enu} from $\nu_e\nu_e$.
It lowers the production of the vector bosons by $5$ to $16\%$, which we neglect, so that their signals are slightly overestimated.

We describe the core by one zone of $9000\,\mathrm{km}^3$ that radiates for $3\,\mathrm{s}$~\cite{Caputo:2022mah}.
We consider two supernova models, a cold core with $T=30$ to $40\,\MeV$ and a hot one with $T=50$ to $60\,\MeV$. The cold core is close to the coldest Garching model used in recent bounds~\cite{Caputo:2021rux} and the hot one to the hottest; we count a particle as excluded only if the cold core excludes it.
Both have electron chemical potentials of $100$ to $200\,\MeV$, and for the HNL and the scalar, electron-neutrino chemical potentials of $100$ to $150\,\MeV$.

A single zone does not reproduce the emission of a realistic core, so we calibrate it.
For each zone, we compute the SMM signal of an axion-like particle coupled to electrons, at the same mass and at the edge of its SN~1987A bound~\cite{Fiorillo:2025sln}, and divide all predicted signals by the ratio of this signal to the SMM limit.

\paragraph{Signals and their limits.}
The particles leave the core and decay about $10^4\,\mathrm{s}$ later in their rest frame, mostly far outside the progenitor.
Their decays into $e^+e^-$ radiate photons, the HNL emits one in each decay, and the positrons slow down in the interstellar gas.
We compare these products with the following limits:
\begin{itemize}
\item photons of $25$ to $100\,\MeV$ that reach the Earth within the $223\,\mathrm{s}$ observed by SMM after SN~1987A, below $1.84\,\mathrm{cm}^{-2}$ as prescribed in Ref.~\cite{Fiorillo:2025sln};
\item photons of $0.1$ to $3\,\MeV$ recorded by PVO in the $8000\,\mathrm{s}$ after the burst, compared with the limit of Ref.~\cite{Jaffe:1995sw}, derived for its template spectrum and time profile, and, in the $0.5$ to $1\,\MeV$ channel, with a three-standard-deviation excess over its background of $10$ counts per second~\cite{Chauhan:2025xqr}, for an effective area of $17\,\mathrm{cm}^2$;
\item positrons released outside the progenitor, at most $1.4\times10^{52}$ per supernova by the Galactic $511\,\keV$ line~\cite{Calore:2021lih}, and the gamma rays from their annihilation in flight~\cite{DelaTorreLuque:2024zsr};
\item energy deposited inside the progenitor, at most $10^{50}\,\mathrm{erg}$ by low-energy supernovae~\cite{Caputo:2022mah};
\item photons from the neutron-star merger GW170817, for which we check that the injected pairs remain optically thin and compare the photons that escape the merger ejecta with the Fermi-LAT limit, as in Ref.~\cite{Diamond:2023cto}.
\end{itemize}
For the vector bosons, the $511\,\keV$ photons of positrons that annihilate in the envelope, the signal analyzed in Ref.~\cite{Chauhan:2025xqr}, stay three to five times below its sensitivity.

\paragraph{A Galactic supernova.}
A supernova at $10\,\mathrm{kpc}$ would test these particles far below the SN~1987A limits~\cite{Muller:2023vjm}.
With the same production, we count the photons above $30\,\MeV$ that the Fermi Large Area Telescope detects within $10^4\,\mathrm{s}$ after the neutrino burst while the supernova stays in its field of view, for effective areas of $500$, $2500$, and $6000\,\mathrm{cm}^2$ from $30$ to $100\,\MeV$, $100$ to $300\,\MeV$, and $0.3$ to $1\,\mathrm{GeV}$~\cite{Fermi-LAT:2021wbg}.
For cold to hot cores and red to blue supergiant progenitors, the vector bosons give $10^4$ to $10^6$ photons, and the HNL, which emits a photon in each decay, $4\times10^6$ to $3\times10^8$.
The survey views a given direction about a fifth of the time, which leaves at least $3\times10^3$ photons, against about $10^3$ from the diffuse emission of the inner Galaxy~\cite{Fermi-LAT:2012edv} within the angular resolution of the telescope.

\paragraph{Decays through lighter bosons.}
A parent of mass $m_X\approx2m$ that decays at rest into two of these bosons gives the same electrons and neutrinos with its own lifetime, which separates the lifetime from the couplings of the bosons to the Standard Model.
Supernovae then make the parent only through its own coupling and with the factor $e^{-m_X/T}$, and the signals of the bosons follow from their couplings and decay lengths, not from the lifetime.
Faster bosons, from a heavier parent, spread the energies of the electrons and neutrinos up to $m_X/2$, and neutrinos above about $180\,\MeV$ make more $^3$He.

\paragraph{Q-balls.}
Supernovae cannot produce the Q-balls of the lithium solutions: at their peak charges, populations 1 to 3 weigh $(0.2$ to $1)\times10^{26}\,\mathrm{GeV}$, and even their constituents, the sleptons of the supersymmetric model of Sec.~\ref{sec:embedding}, weigh hundreds of GeV.

\section{Q-ball charge distribution and emission}\label{sec:source}
In this section, we specify the evaporation of a single Q-ball, the distribution of initial charges in a population, and the composition of the emitted particles.

As in the main text, a Q-ball is a localized configuration of a complex scalar field with the lowest energy at fixed conserved charge $Q$~\cite{Coleman:1985ki}.
When the scalar potential is nearly constant, at a value $V_0$, over the field values inside it, its mass $M_Q$ and the mean energy $\omega_Q$ of an emitted electron or neutrino are~\cite{Kasuya:2025nix}
\begin{equation}
 M_Q=\frac{4}{3}CQ^{3/4},\qquad
 \omega_Q=\frac{dM_Q}{dQ}=CQ^{-1/4},
 \label{eq:S-mass}
\end{equation}
where $C$ combines $V_0^{1/4}$ with a numerical factor from the field profile.
These Q-balls evaporate at their surface, where pairs of their quanta turn into pairs of electrons or neutrinos, one per unit of charge~\cite{Cohen:1986ct}.
Pauli blocking of the outgoing fermions limits the emission rate per unit area to a value proportional to $\omega_Q^3$~\cite{Cohen:1986ct}; a Q-ball that emits at this maximal rate loses charge as $\dot Q=-AQ^{-1/4}$, with the surface coefficient $A=\pi C/(3\hbar)$~\cite{Kasuya:2025nix}.
A soliton with initial charge $Q_0$ therefore evolves as
\begin{equation}
 \tau_Q=\frac{4Q_0^{5/4}}{5A},\qquad
 \frac{Q(t)}{Q_0}=\left(1-\frac{t}{\tau_Q}\right)^{4/5},
 \label{eq:S-evolution}
\end{equation}
where $t$ is its age and $\tau_Q$ its lifetime; the energy of its emitted particles grows as $\omega_Q\propto(1-t/\tau_Q)^{-1/5}$.

Fragmentation of the scalar condensate produces Q-balls with a distribution of initial charges.
With $x=Q_0/Q_{\rm peak}$, where $Q_{\rm peak}$ is the charge at the maximum of this distribution, the initial energy density of the Q-balls, $\rho_{Q,0}$, is distributed as
\begin{equation}
 \frac{d\rho_{Q,0}}{d\ln x}=\rho_{Q,0}W(x),\qquad
 W(x)=\mathcal N x^{1/2}e^{-x^2/4},
 \label{eq:S-distribution}
\end{equation}
where $\mathcal N$ normalizes $\int W\,d\ln x=1$ over the range of $x$ given below.
The solitons with $Q_0=Q_{\rm peak}$ define the peak lifetime \tpeak and the peak energy $\omega_{\rm peak}$ of the emitted particles at $t=0$, so that $\tau_Q=\tpeak x^{5/4}$ and $\omega_Q(x,0)=\omega_{\rm peak}x^{-1/4}$.
For given $\omega_{\rm peak}$ and \tpeak, Eqs.~\eqref{eq:S-mass} and \eqref{eq:S-evolution} at $Q_0=Q_{\rm peak}$ fix $C$, $A$, and $Q_{\rm peak}$.
Since $\tau_Q$ grows with the initial charge, the charge distribution fixes both the early neutrino emission and how long electromagnetic energy enters the plasma after nuclear burning.

Ref.~\cite{Kasuya:2025nix} simulated the fragmentation of the condensate on a lattice in twelve runs and fitted the resulting charge distribution; in our variables, this fit is Eq.~\eqref{eq:S-distribution}.
The simulated Q-balls cover $x\geq0.04$, and ten of the runs reach $x=10^{-2}$.
Between $x=10^{-2}$ and $0.5$, the simulated $W$ follows Eq.~\eqref{eq:S-distribution} with an rms scatter of $0.2$ in $\log_{10}W$ and rises as $x^{0.4\pm0.2}$.
We apply Eq.~\eqref{eq:S-distribution} over $10^{-6}\leq x\leq8$.
For the lithium solutions, emission after $200\,\mathrm{s}$, while $^7$Be forms and deuterium is photodissociated, comes only from solitons with $x\geq(200\,\mathrm{s}/\tpeak)^{4/5}=0.02$ to $0.04$, inside the simulated charges.
Below $x=10^{-2}$, where the runs differ by factors of three or more, Eq.~\eqref{eq:S-distribution} places $8\%$ of the energy, which the lithium solutions release within about $80\,\mathrm{s}$; Sec.~\ref{sec:variations} varies this part of the distribution.

A population contains Q-balls of quanta and of antiquanta, which follow the same distribution and carry the energy fractions $f_\pm=(1\pm\epsilon)/2$.
In electron populations, Q-balls of quanta emit electrons and neutrinos in the numbers of Eq.~\eqref{eq:S-composition}~\cite{Kasuya:2025nix}: a third are $\nu_\mu$, a fraction $b$ of the others are $e^-$ and $e^+$, and the rest are $\nu_e$ and $\bar\nu_e$.
In muon populations, which we use for comparison, a third are $\bar\nu_\tau$, a fraction $b$ of the others are $\mu^+$ and $e^-$, and the rest are $\nu_e$ and $\bar\nu_\mu$.
Q-balls of antiquanta emit the conjugates; in the populations of Table~\ref{tab:soft-points}, they carry all the energy for $\epsilon=-1$ and three quarters of it for $\epsilon=-0.5$.

The emitted particles divide equally the mass that the Q-ball loses: when its charge decreases from $Q_a$ to $Q_b$, each of the $Q_a-Q_b$ particles carries
\begin{equation}
 \bar\omega=\frac{M_Q(Q_a)-M_Q(Q_b)}{Q_a-Q_b}.
 \label{eq:finite-energy}
\end{equation}
This prescription conserves the Q-ball energy and approaches $\omega_Q(Q)$ as $Q_a-Q_b\to0$.
The particles are created in pairs, so their energies spread around this mean~\cite{Cohen:1986ct,Kawasaki:2012gk}; Sec.~\ref{sec:variations} shows the effect of such a spread.

Charged emission opens at $\omega_Q=m_e$ for the electron population and at $m_\mu$ for the muon populations.
Near $Q=0$, $\omega_Q$ reaches higher thresholds, first those of channels with muons (Sec.~\ref{sec:variations}), and the potential departs from its plateau, where Eqs.~\eqref{eq:S-mass} and \eqref{eq:S-evolution} cease to apply.

We represent the charge distribution of Eq.~\eqref{eq:S-distribution} by discrete charges, each standing for many identical solitons that evolve according to Eq.~\eqref{eq:S-evolution}.
Their charge loss fixes the number of emitted particles, and Eq.~\eqref{eq:finite-energy} divides the released mass among the species of Eq.~\eqref{eq:S-composition}; as the charge steps and the spacing of the discrete charges decrease, the emission approaches its continuous limit (Sec.~\ref{sec:numerics}).

The transport calculation starts at the photon temperature $T_0=3\,\MeV$, at the age $t_0$ of the Universe.
Solitons formed earlier enter with charges reduced by Eq.~\eqref{eq:S-evolution}, and those with $\tau_Q\leq t_0$ have disappeared; $R_Q$ denotes $\rho_Q/\rho_{\rm rad}$ at $T_\gamma=3\,\MeV$.
For $\tpeak\leq0.3\,\mathrm{s}$, the emission between $5$ and $3\,\MeV$ is comparable to $R_Q\rho_{\rm rad}$ and thermalizes only partly, so these populations start at $T_0=5\,\MeV$.
Even without the source, this choice shifts \yp by up to $10^{-4}$; we subtract the offset, so that all lifetimes compare with the same standard BBN.
For longer lifetimes, the initial temperature of $3\,\MeV$ moves the boundaries of the allowed regions by $2\%$ or less.

We take the emission before $T_0$ to be thermalized.
Its lepton number, one third per unit of charge, enters the spectra at $T_0$ as Fermi-Dirac chemical potentials $\mu_k=\xi_kT_0$ of the three propagation eigenstates, with the thermal mixing of Sec.~\ref{sec:transport}.
The neutron fraction then starts at $[1+\exp(\Delta_{np}/T_0+\xi_e)]^{-1}$, with $\Delta_{np}=m_n-m_p$ and $\xi_e=\sum_k|U_{ek}|^2\xi_k$.

\section{How Q-ball populations and single particles lower lithium-7}\label{sec:emission}
A Q-ball population and a decaying species with one lifetime differ in when they release their energy and in the energies of the electrons and neutrinos they emit.
In this section, we show how these differences change the abundances.

\paragraph{Comparison with one species.}
For one species, the neutrons from its antineutrinos raise \dh, while its cascades lower \dh only once $E_c$ exceeds the deuteron threshold, at $T_\gamma\simeq5\,\keV$.
Both effects scale with the number of decays, so the measured \dh fixes the lifetime near $10^4\,\mathrm{s}$, almost independently of the abundance.
Most decays then occur after ${}^7\mathrm{Li}(p,\alpha){}^4\mathrm{He}$ stops, near $3\times10^3\,\mathrm{s}$.
A neutron captured on $^7$Be then turns it into $^7$Li, which survives, and photons break up this $^7$Li only above $2.47\,\MeV$, higher than the $1.59\,\MeV$ that suffice for $^7$Be.

To lower $^7$Li enough, the species therefore needs more decays or more energetic antineutrinos.
More decays raise \yp, through the expansion rate and the higher $\eta_B$ during nucleosynthesis; their effect on \neff depends on how their energy divides between neutrinos and the electromagnetic plasma.
More energetic antineutrinos convert more protons per unit of energy but break up more $^4$He into the fragments that make $^3$He and $^6$Li.
A species that emits the electrons and neutrinos of population 3 with one lifetime, at either their peak or their power-weighted mean energy, lowers $^7$Li only to $4.1$ or $3.0\times10^{-10}$, respectively, against $2.0\times10^{-10}$ for the population, before \yp reaches its upper limit (Fig.~\ref{fig:single-map}).

\begin{figure}[t!]
\centering
\includegraphics[width=0.49\textwidth]{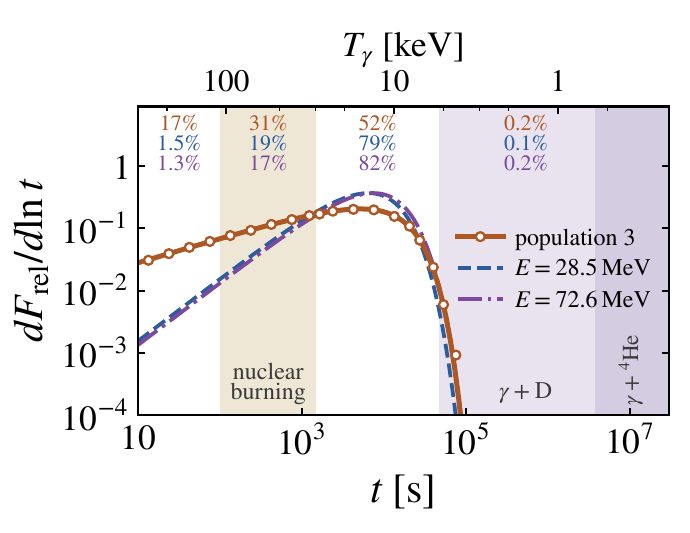}\hfill
\includegraphics[width=0.49\textwidth]{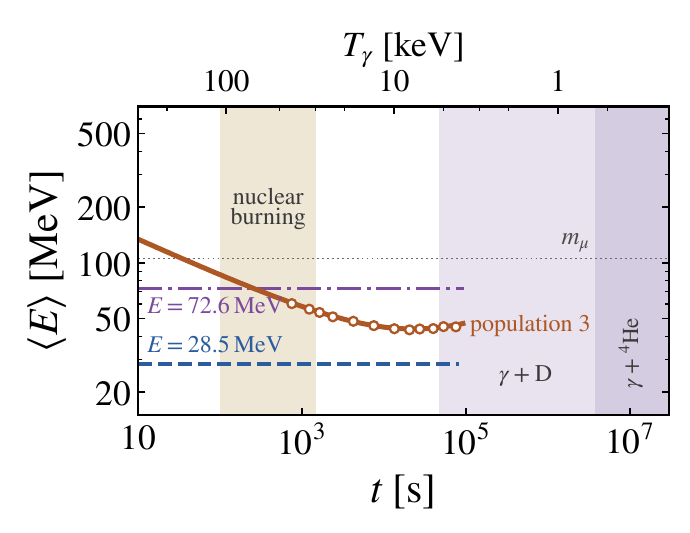}
\caption{Energy injection by Q-ball population 3 and by one decaying species with the same emitted particles, each at the lifetime at which $10^5\dh=2.508$.
Population 3: orange, with circles from its transport calculations; one species $X$ with a single lifetime and a single energy $E=28.5$ or $72.6\,\MeV$: blue dashed and purple dash-dotted, with the abundance that gives the \neff of population 3.
Left: energy released per logarithmic interval of time, as a fraction of the source energy at $T_0$; numbers give the fractions released in the four intervals.
Right: power-weighted mean energy of the emitted electrons and neutrinos.
Shading marks nuclear burning and the temperatures at which $E_c$ exceeds the thresholds of D and $^4$He breakup.}
\label{fig:emission-history}
\end{figure}

A Q-ball population instead emits in two stages (Fig.~\ref{fig:emission-history}).
Its small solitons emit energetic electrons and neutrinos early, and its large solitons less energetic ones later, with abundances and lifetimes that the charge distribution fixes together.
A larger fraction of its energy therefore arrives while neutrons still lower $^7$Li, so the population needs less energy than one species and shifts \yp less.
Its late emission, which destroys the extra deuterium, is softer and breaks up less $^4$He, which lowers $^6\mathrm{Li}/{}^7\mathrm{Li}$ and $^3$He.

The value of \neff follows in both cases from the energies given to neutrinos and to photons, so particles and Q-ball populations that are lithium solutions span similar ranges of it (Table~\ref{tab:solutions}).
Stars destroy $^6$Li more readily than $^7$Li, by factors that stellar models predict only within wide ranges~\cite{Asplund:2005yt}; the lower primordial $^6\mathrm{Li}/{}^7\mathrm{Li}$ of Q-ball populations with $\omega_{\rm peak}\lesssim50\,\MeV$ requires less of this depletion to meet the stellar limits.

\begin{figure*}[t!]
\centering
\includegraphics[width=0.49\textwidth]{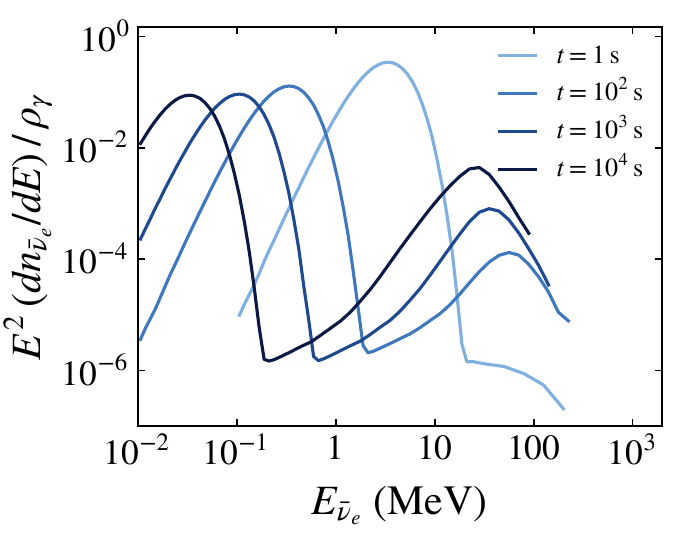}\hfill
\includegraphics[width=0.49\textwidth]{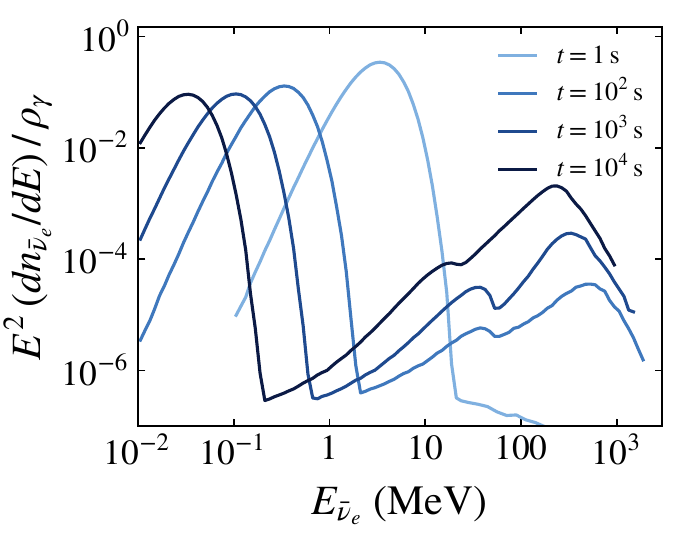}
\caption{Spectra of electron antineutrinos at four ages $t$ of the Universe, as $E^2\,dn_{\bar\nu_e}/dE$ divided by the photon energy density $\rho_\gamma=\pi^2T_\gamma^4/15$, where $dn_{\bar\nu_e}/dE$ is their number density per unit energy $E$.
Left: population 3, an electron population with $\omega_{\rm peak}=28.5\,\MeV$, $\tpeak=1.04\times10^4\,\mathrm{s}$, $b=1$, $\epsilon=-1$, and $R_Q=5.0\times10^{-4}$.
Right: muon population with $\omega_{\rm peak}=238\,\MeV$, $\tpeak=1.63\times10^4\,\mathrm{s}$, $b=1$, $\epsilon=+1$, and $R_Q=2\times10^{-4}$.}
\label{fig:nu-spectra}
\end{figure*}

\paragraph{Antineutrino spectra.}
In a population, the solitons with small charges evaporate first and emit the most energetic electrons and neutrinos, and those with large charges emit less energetic ones hours later.
Fig.~\ref{fig:nu-spectra} shows the resulting spectra of electron antineutrinos at four times for population 3 and for a muon population.
At every epoch, the thermal distribution peaks near $4T_\nu$ and falls exponentially above it.
Above about $20T_\nu$, the antineutrinos from the solitons dominate.
Each has redshifted since its emission, so the spectrum rises toward a peak set by the solitons emitting at that time.
At $1\,\mathrm{s}$, few solitons have evaporated, and the emitted antineutrinos form a faint tail.
At $10^2\,\mathrm{s}$, small solitons in their final evaporation place the peak near $2\,\omega_{\rm peak}$; by $10^4\,\mathrm{s}$, solitons with charges near $Q_{\rm peak}$ dominate, and the peak lies at $\omega_{\rm peak}$.

In the muon population, the solitons emit $\bar\nu_\tau$, which oscillations partly turn into $\bar\nu_e$.
Until about $2\times10^3\,\mathrm{s}$, the plasma stops the muons before they decay, and their decays at rest add antineutrinos up to $m_\mu/2=52.8\,\MeV$, which produce the step at $10^2$ and $10^3\,\mathrm{s}$.
Later muons decay in flight, and the step from earlier decays redshifts into the shoulder at lower energies.

\begin{figure}[t!]
\centering
\includegraphics[width=0.49\textwidth]{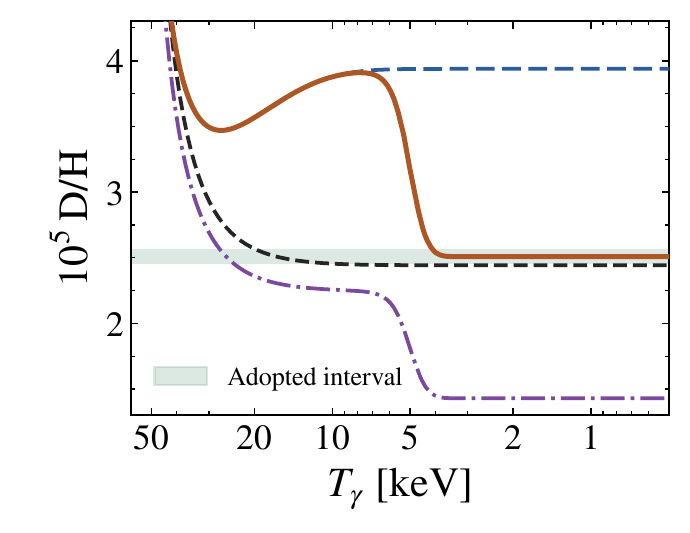}\hfill
\includegraphics[width=0.49\textwidth]{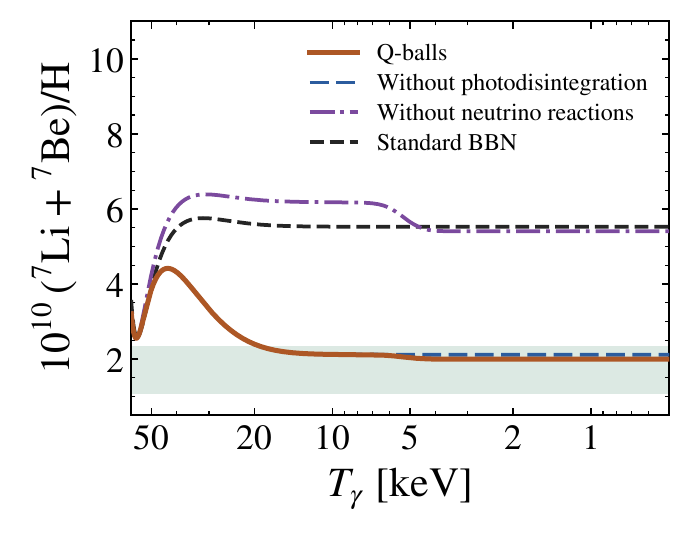}
\caption{\dh (left) and $(^7\mathrm{Li}+{}^7\mathrm{Be})/\mathrm{H}$ (right) versus photon temperature for population 3 (solid), recalculated without photodisintegration (blue dashed) and without the reactions of the injected neutrinos with nucleons and nuclei (purple dash-dotted); black dashed curves give standard BBN.
All calculations end with $\eta_B^{\rm CMB}=6.13\times10^{-10}$~\cite{Planck:2018vyg} and average four independent simulations at the peak lifetime where the population yields $10^5\dh=2.508$, as in Table~\ref{tab:soft-points} (Sec.~\ref{sec:map}).
Green shading: \dh interval of Eq.~\eqref{eq:S-abundances}~\cite{ParticleDataGroup:2026bbn} and stellar $^7$Li band~\cite{Sbordone:2010zi,Charbonnel:2026lip}.}
\label{fig:mechanism}
\end{figure}

\paragraph{Neutron capture and photodisintegration.}
Fig.~\ref{fig:mechanism} separates the neutron capture and photodisintegration that set \dh for population 3.
Without photodisintegration, \dh would end $61\%$ above standard BBN, and without the reactions of the injected neutrinos, $41\%$ below; together they give the measured \dh and lower $^7$Li by about two thirds, almost entirely through the neutrons.
The calculation without the neutrino reactions also isolates the photon heating.
Energy released after nuclear burning raises $\eta_B$ by about $6\%$ while nuclei form, which lowers \dh and raises $^7$Li before photodisintegration begins.

\section{Q-ball populations allowed by the measurements}\label{sec:lifetimes}
In this section, we map the Q-ball populations allowed by the measurements and test the robustness of the lithium solutions.

\begin{table}[t!]
\centering
\footnotesize
\begin{tabular*}{\columnwidth}{@{}l@{\hspace{4pt}\extracolsep{\fill}}lll@{}}
\toprule
Symbol & Meaning & Computed & \begin{tabular}[b]{@{}l@{}}Lithium\\solutions\end{tabular}\\
\midrule
$\omega_{\rm peak}$ & energy at $Q_{\rm peak}$ & $0.025$ to $2.6\,\mathrm{GeV}$ & $28.5$ to $120\,\MeV$\\
\tpeak & lifetime at $Q_{\rm peak}$ & $0.1$ to $10^7\,\mathrm{s}$ & $(1$ to $2.1)\times10^4\,\mathrm{s}$\\
$R_Q$ & $\rho_Q/\rho_{\rm rad}$ at $3\,\MeV$ & $10^{-10}$ to $0.1$ & $(0.6$ to $6.4)\times10^{-4}$\\
$b$ & charged fraction & $0.25$ to $1$ & $0.25$ to $1$\\
$\epsilon$ & charge asymmetry & $-1$ to $1$ & $-1$ to $0$\\
\bottomrule
\end{tabular*}
\caption{Parameters of a population of leptonic Q-balls, whose charges follow Eq.~\eqref{eq:S-distribution}; $\omega_{\rm peak}$ is the energy of the electrons and neutrinos emitted at $Q_{\rm peak}$, and $b$ is the fraction of the emitted particles other than $\nu_\mu$ that are $e^\pm$, which thus receive $2b/3$ of the energy.
The last column gives the ranges of the lithium solutions, Eq.~\eqref{eq:S-lithium}, among electron populations with $28.5\leq\omega_{\rm peak}\leq120\,\MeV$, $b\geq0.25$, and $\epsilon\leq0$; they reach every edge of this range.}
\label{tab:parameters}
\end{table}

\begin{figure}[t!]
\centering
\includegraphics[width=0.6\textwidth]{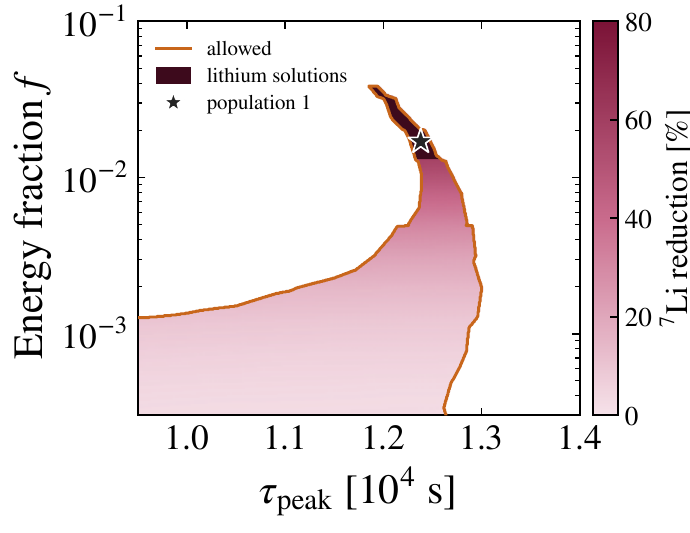}
\caption{Largest energy fraction $f$, Eq.~\eqref{eq:f}, allowed by Eq.~\eqref{eq:S-abundances} versus the peak lifetime near $10^4\,\mathrm{s}$, for populations with the $\omega_{\rm peak}$, $b$, and $\epsilon$ of population 1; color gives the $^7$Li reduction relative to standard BBN, and dark red marks lithium solutions, Eq.~\eqref{eq:S-lithium}; star: population 1.}
\label{fig:region}
\end{figure}

A population is allowed when it satisfies Eq.~\eqref{eq:S-abundances}.
Standard BBN gives $10^5\dh=2.443$, just below the measured interval, so an allowed population must raise \dh.
At each peak lifetime, the allowed values of the energy fraction, the largest $\rho_Q/\rho_{\rm rad}$ in the standard expansion history, therefore lie between a lower edge, where $10^5\dh$ reaches $2.450$, and an upper edge, where \dh or another observable leaves its interval.
For the class of population 1, with $\omega_{\rm peak}=41\,\MeV$, $b=0.75$, and $\epsilon=-1$, allowed populations exist for peak lifetimes from $0.3$ to $5\times10^4\,\mathrm{s}$, and none at longer lifetimes up to $10^7\,\mathrm{s}$; an electron population with $b\approx0.3$ and muon populations give allowed lifetimes from about $0.4$ to $3\times10^4\,\mathrm{s}$.
Fig.~\ref{fig:region} shows the peak of these regions near $10^4\,\mathrm{s}$. Their shape follows from what the emitted neutrinos, neutrons, and photons do at each epoch.

Below $\tpeak\approx1\,\mathrm{s}$, much of the emission occurs before neutrinos decouple and leaves excess muon antineutrinos, which oscillations partly turn into electron antineutrinos that raise \yp, so \yp leaves its interval before $10^5\dh$ reaches $2.450$.
For lifetimes of a few seconds, the neutrons produced by energetic neutrinos end in helium, and \yp bounds the energy fraction; from tens of seconds to a few hours, they are captured into deuterium, and \dh bounds it.
Outside the peak, $^7$Li changes by at most about $10\%$.

Near $10^4\,\mathrm{s}$, neutron capture raises \dh while photodisintegration lowers it, and where the two cancel, a lifetime with \dh in its interval exists at every energy fraction until another constraint fails.
These cancelling lifetimes form the narrow peak of Fig.~\ref{fig:region}, and Fig.~\ref{fig:mechanism} separates the two processes for population 3.

The peak of Fig.~\ref{fig:region} ends where \yp reaches its upper limit, at an energy fraction of about $4\%$, where $^7$Li approaches the lower edge of the stellar band.
Lithium solutions cover its upper part, from about $1.3$ to $3.8\%$, at peak lifetimes of $1.19$ to $1.26\times10^4\,\mathrm{s}$.
Four independent simulations of populations with $b=0.75$ and $\epsilon=-1$ at $f=3$ and $3.5\%$, the larger with \yp within $2\times10^{-4}$ of its limit, give $\li=1.03$ to $1.30\times10^{-10}$ for $\omega_{\rm peak}$ from $28.5$ to $90\,\MeV$, lowest near $55\,\MeV$, with $^6\mathrm{Li}/{}^7\mathrm{Li}$ rising from $0.5$ to $1\%$ at $28.5\,\MeV$ to $18$ to $31\%$ at $90\,\MeV$; at $105$ and $120\,\MeV$, \hethree hits its bound first (upper panel of Fig.~\ref{fig:qball-abundances} of the main text).
The populations keep emitting after their energy fraction reaches its maximum and release in total about $2.4$ times this maximum for $\tpeak\sim10^4\,\mathrm{s}$.
Beyond the peak, photodisintegration only lowers \dh, and at $\tpeak=10^7\,\mathrm{s}$, breakup of $^4$He raises \hethree to six times its bound.

The regions come from transport calculations of this class with $N_\nu=10^6$, on a grid of peak lifetimes from $0.5$ to $10^7\,\mathrm{s}$ and energy fractions that is densest around the peak.

Between them, we interpolate linearly in $\log\tpeak$ and in the logarithm of the energy fraction; below $2\times10^3\,\mathrm{s}$, we smooth the boundaries of the regions over $0.06$ in $\log\tpeak$ and $0.03$ in the logarithm of the energy fraction, which preserves the allowed or excluded classification of every computed point.
For \yp, whose scatter of $2.6\times10^{-4}$ per calculation is comparable to its distance from the upper limit, we use one smooth fit, $\yp=c_0+fP_5(\log\tpeak)+f^2Q_5(\log\tpeak)$ in the energy fraction $f$ with polynomials $P_5$ and $Q_5$ of fifth degree, which has $\chi^2$ per degree of freedom $1.00$.

\begin{figure}[t!]
\centering
\includegraphics[width=0.6\textwidth]{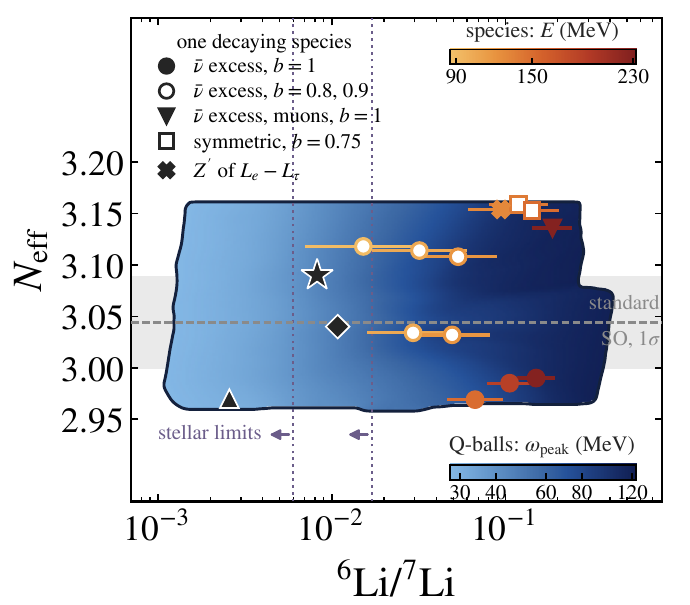}
\caption{$^6\mathrm{Li}/{}^7\mathrm{Li}$ and \neff of lithium solutions, Eq.~\eqref{eq:S-lithium}, with \dh at its measured value.
Blue region: Q-ball populations with $\omega_{\rm peak}$, $b$, $\epsilon$, and $R_Q$ across the ranges of Table~\ref{tab:parameters}, shaded by the geometric mean of $\omega_{\rm peak}$ at each point; star, diamond, and triangle: populations 1 to 3.
Markers with bars: one decaying species of Sec.~\ref{sec:single-hard}, colored by the energy $E$ of each emitted particle, and the gauge boson $Z'$ of $L_e-L_\tau$.
The bars and the region span two recoil prescriptions for helium fragments.
Gray band: $1\sigma$ precision forecast for the Simons Observatory (SO)~\cite{SimonsObservatory:2025wwn} around the standard \neff (dashed line); dotted lines: smallest and largest $2\sigma$ upper limits on $^6\mathrm{Li}/{}^7\mathrm{Li}$ in metal-poor stars~\cite{Wang:2021urb}.}
\label{fig:family}
\end{figure}

\subsection{Map of Q-ball populations that solve the lithium problem}\label{sec:map}
Figure~\ref{fig:family} shows the values of \neff and $^6\mathrm{Li}/{}^7\mathrm{Li}$ that electron populations reach as lithium solutions, Eq.~\eqref{eq:S-lithium}, with \dh at its measured value.
Since the region ends where \yp reaches its upper limit, helium determines its edge at low \neff.
A larger $R_Q$ lowers \li, \hethree, and \neff and raises \yp and $^6\mathrm{Li}/{}^7\mathrm{Li}$; along the three rows of Table~\ref{tab:soft-points} with $\omega_{\rm peak}=28.5\,\MeV$, \yp reaches its upper limit at $R_Q\approx5.6\times10^{-4}$, where $\neff=2.965$ and $\li\approx1.8\times10^{-10}$.
A smaller $b$ raises \dh at fixed \tpeak, since less charge then leaves as $e^+e^-$ and more as $\bar\nu_e\nu_e$, whose antineutrinos convert protons into neutrons, so the lifetime at which \dh equals its measured value lengthens by $0.5$ to $0.8\%$ per $0.01$ in $b$.

We compute with full transport populations with $\omega_{\rm peak}$ from $28.5$ to $120\,\MeV$ and $b$ from $0.25$ to $1$, with $\epsilon=-1$ and $-0.5$ and two to six values of $R_Q$ for each.
Further asymmetries cover $\epsilon=-0.75$ for $b\lesssim0.5$, and $\epsilon=-0.25$ and $0$ for $\omega_{\rm peak}\leq45\,\MeV$ and $0.85\leq b\leq0.95$.
Each population is computed at two or three peak lifetimes with $N_\nu=10^6$, and straight-line fits in \tpeak give its abundances and \neff where \dh equals its measured value.

Ten populations, including populations 1 to 3, are also computed in four independent simulations with $N_\nu=2\times10^6$ (Table~\ref{tab:soft-points}).
Each simulation computes the population at three peak lifetimes a few percent apart, which bracket the one at which $10^5\dh=2.508$, and least-squares lines in \tpeak give every quantity there; one line through all twelve calculations of a population, or a parabola through each simulation, changes the means by at most the scatter of one calculation.
Since \dh falls by $2$ to $3\%$ per percent increase of \tpeak, the error of each mean corresponds to at most $0.3\%$ in \dh, a seventh of the half-width of its interval.

\begin{table}[t!]
\centering
\footnotesize
\setlength{\tabcolsep}{2.8pt}
\begin{tabular}{@{}lccccccccccc@{}}
\toprule
Population & $\omega_{\rm peak}$ [MeV] & $b$ & $\epsilon$ & $10^4R_Q$ & \tpeak [$10^4$\,s] & $10^5\dh$ & \yp & $10^6\,{}^3\mathrm{He/H}$ & $10^{10}\,{}^7\mathrm{Li/H}$ & \neff & $10^2\,{}^6\mathrm{Li}/{}^7\mathrm{Li}$\\
\midrule
& $28.5$ & $1$ & $-1$ & $5.8$ & $1.038(2)$ & $2.508$ & $0.2485(1)$ & $8.513(4)$ & $1.771(6)$ & $2.961(1)$ & $0.225(1)$ to $0.506(2)$\\
& $28.5$ & $1$ & $-1$ & $5.3$ & $1.040(1)$ & $2.508$ & $0.2483(2)$ & $8.628(3)$ & $1.903(8)$ & $2.970(1)$ & $0.194(1)$ to $0.434(2)$\\
3 & $28.5$ & $1$ & $-1$ & $5.0$ & $1.041(1)$ & $2.508$ & $0.2482(1)$ & $8.699(4)$ & $1.997(9)$ & $2.972(1)$ & $0.174(1)$ to $0.392(2)$\\
\midrule
1 & $41$ & $0.75$ & $-1$ & $2.39$ & $1.238(2)$ & $2.508$ & $0.2476(1)$ & $9.224(6)$ & $1.937(4)$ & $3.089(2)$ & $0.573(7)$ to $1.195(12)$\\
& $41$ & $0.9$ & $-0.5$ & $3.11$ & $1.113(2)$ & $2.508$ & $0.2478(1)$ & $9.459(6)$ & $2.271(6)$ & $3.040(2)$ & $0.440(6)$ to $0.932(9)$\\
& $45$ & $0.85$ & $-0.5$ & $2.71$ & $1.158(1)$ & $2.508$ & $0.2477(1)$ & $9.611(8)$ & $2.192(4)$ & $3.057(1)$ & $0.609(3)$ to $1.271(4)$\\
2 & $55$ & $0.9$ & $-1$ & $2.0$ & $1.208(3)$ & $2.508$ & $0.2476(1)$ & $9.500(4)$ & $2.183(2)$ & $3.040(2)$ & $0.764(7)$ to $1.537(9)$\\
& $60$ & $0.9$ & $-0.5$ & $2.31$ & $1.177(1)$ & $2.508$ & $0.2475(1)$ & $10.090(5)$ & $2.214(4)$ & $3.040(1)$ & $1.129(5)$ to $2.236(8)$\\
\midrule
& $45$ & $0.31$ & $-1$ & $1.15$ & $1.654(2)$ & $2.508$ & $0.2472(2)$ & $9.781(1)$ & $2.281(1)$ & $3.154(1)$ & $0.654(3)$ to $1.332(4)$\\
& $60$ & $0.31$ & $-1$ & $1.0$ & $1.733(3)$ & $2.508$ & $0.2471(1)$ & $10.311(3)$ & $2.177(2)$ & $3.144(1)$ & $1.444(12)$ to $2.793(18)$\\
\midrule
3, $\mu$ & $28.5$ & $1$ & $-1$ & $5.0$ & $1.047(1)$ & $2.508$ & $0.2483(1)$ & $8.655(3)$ & $1.920(9)$ & $2.977(1)$ & $0.190(2)$ to $0.410(3)$\\
2, $\mu$ & $55$ & $0.9$ & $-1$ & $2.0$ & $1.234(1)$ & $2.508$ & $0.2474(2)$ & $9.378(4)$ & $2.026(10)$ & $3.051(1)$ & $0.881(3)$ to $1.715(4)$\\
\midrule
SBBN &  &  &  &  &  & $2.443(3)$ & $0.2471(3)$ & $10.398(4)$ & $5.532(3)$ & $3.045(4)$ & \\
\bottomrule
\end{tabular}
\caption{Light-element abundances, \neff, and $^6\mathrm{Li}/{}^7\mathrm{Li}$ of electron populations with the charge distribution of Eq.~\eqref{eq:S-distribution}, each at the peak lifetime at which $10^5\dh=2.508$, and in standard BBN (SBBN).
Entries give the mean over four independent simulations and, in parentheses, the standard deviation in units of the last digit; each simulation computes the population at three peak lifetimes with $2\times10^6$ simulated neutrinos per calculation, and the SBBN row averages four calculations without Q-balls with the same number of neutrinos.
The rows labelled 1, 2, and 3 are populations 1 to 3 of Table~\ref{tab:solutions}, and the three rows with $\omega_{\rm peak}=28.5\,\MeV$ differ only in $R_Q$; rows $\mu$ add to populations 2 and 3 the $\mu^-\nu_e$ pairs of the supersymmetric model of Sec.~\ref{sec:embedding} (Sec.~\ref{sec:variations}).
${}^3\mathrm{He/H}$ includes surviving tritium, $^7\mathrm{Li/H}$ surviving $^7$Be, and the isotope ratios span the two recoil prescriptions of Sec.~\ref{sec:fragments}.
All populations except the one with $R_Q=5.8\times10^{-4}$ are lithium solutions, Eq.~\eqref{eq:S-lithium}; that one has \yp slightly above the upper limit of Eq.~\eqref{eq:S-abundances}, by twice the standard error of its mean.}
\label{tab:soft-points}
\end{table}

One calculation with $N_\nu=10^6$ scatters by $2.6\times10^{-4}$ in \yp, about half as much as \yp changes between neighboring computed energy fractions near its upper limit.
We therefore locate this limit with one weighted fit to all populations whose \dh reaches its measured value near their computed lifetimes and to those computed four times with $N_\nu=2\times10^6$,
\begin{equation}
 \yp=c_0+\frac{R_Q}{3\times10^{-4}}\left[a_0+a_1\ln\frac{\omega_{\rm peak}}{28.5\,\MeV}+a_2(b-1)+a_3(\epsilon+1)+a_4(b-1)(\epsilon+1)\right],
 \label{eq:yp-fit}
\end{equation}
with $\chi^2$ per degree of freedom $0.996$.
Fits that leave out one value of $\omega_{\rm peak}$, $b$, or $\epsilon$ predict the omitted populations within their statistical errors, and the fit places the limit for population 3 at $R_Q=5.36\times10^{-4}$, $4\%$ below the four-simulation mean; quadratic or cubic terms move it by at most $5\%$.
Only the helium limit uses this global fit; \dh, the other abundances, and \neff of each population come from its own calculations.

To draw the region, we interpolate linearly in $\ln R_Q$ between the computed energy fractions; below the smallest one we extrapolate by at most one step, and only to locate the upper edge of the stellar $^7$Li band.
Between populations, the interpolation is linear in $\ln\omega_{\rm peak}$, $b$, and $\epsilon$, at the same relative position within their ranges of $R_Q$.
The outline, smoothed over $0.003$ in \neff and a factor of $1.05$ in $^6\mathrm{Li}/{}^7\mathrm{Li}$, contains all computed populations that are lithium solutions, and the shading shows the geometric mean of $\omega_{\rm peak}$ of the populations at each point.
Across the $1\sigma$ interval of \dh, the transport calculations change \li by at most $0.2\%$, $^6\mathrm{Li}/{}^7\mathrm{Li}$ by $0.9\%$, and \neff by $0.001$, so the region found at the central value also represents this range.
The dashed line marks the standard value $\neff=3.044$, which the same calculation without Q-balls reproduces within its statistical error of $0.002$.

\subsection{Robustness of the Q-ball solutions}\label{sec:variations}
The evaporation law, the spectrum and channels of the emitted particles, and the charge distribution are inputs that a microscopic model would fix.
We vary them one at a time and check whether lithium solutions remain.
The solutions depend on how the population divides its emission: energetic antineutrinos within the first hour, while $^7$Be forms, and electromagnetic energy later, which destroys deuterium.
Variations that keep this division keep the solutions, even when they shift the peak lifetime by tens of percent.
Variations that remove part of the small solitons or of the large-charge tail remove the solutions at some of the test points defined below.

We test each variation at three lithium solutions, each taken at the bound that ends its range of $R_Q$ and named by this bound (Table~\ref{tab:variations}).
The \yp point has the $\omega_{\rm peak}$, $b$, and $\epsilon$ of population 3 and ends where \yp reaches its upper limit.
The $^7$Li point, with $\omega_{\rm peak}=41\,\MeV$, $b=0.9$, and $\epsilon=-0.5$, ends where $^7$Li/H reaches the top of the stellar band.
The \neff point, with $\omega_{\rm peak}=60\,\MeV$, $b\approx0.3$, and $\epsilon=-1$, ends where \neff reaches its upper limit.

To locate a bound, we compute the population at two values of $R_Q$ that bracket it.
At each value, three calculations with $N_\nu=2\times10^6$ at nearby peak lifetimes give the lifetime $\tau_*$ at which $10^5\dh=2.508$, from a parabola in \tpeak, and the abundances and \neff there, from straight lines.
Interpolation in $\ln R_Q$ between the two values then gives the bound and the quantities there.
With one simulation per calculation, the bound is uncertain by about $5\%$ in $R_Q$ at the \yp point, where \yp changes by only $0.002$ per unit of $\ln R_Q$, and by $1\%$ at the other two points.
For the reference model, one simulation places the \yp-point bound $12\%$ below the mean of four simulations (Table~\ref{tab:soft-points}); at the other two points, the two agree within $1\%$.
In all variants, \hethree stays below $1.3\times10^{-5}$.

\begin{table}[t!]
\centering
\footnotesize
\setlength{\tabcolsep}{5pt}
\begin{tabular}{@{}lcccc@{}}
\toprule
Variation & $\tau_*/\tau_*^{\rm ref}$ & \yp point & $^7$Li point & \neff point\\
\midrule
$\tau_Q=\tpeak x^{3/2}$ & $0.7$ & yes & yes & yes\\
$\tau_Q=\tpeak x$ & $1.3$ to $1.4$ & yes & yes & yes$^*$\\
\addlinespace[2pt]
$W=0$ for $x<0.01$ & $1.0$ & yes & yes & yes\\
$W=0$ for $x<0.04$ & $1.0$ & yes & yes & yes\\
$W=0$ for $x<0.1$ & $0.9$ to $1.0$ & yes & yes & no\\
\addlinespace[2pt]
$W\propto x^{1/2}e^{-x^2/2}$ & $1.6$ & yes & yes & yes\\
$W\propto x^{1/2}e^{-x^2/8}$ & $0.6$ & yes & yes & yes\\
$W\propto xe^{-x^2/4}$ & $0.8$ & no & marginal & no\\
\addlinespace[2pt]
Lattice runs 1 and 6 & $1.0$; $3.6$ to $4.0$ & yes$^*$ & yes & no\\
Lattice run 2 & $1.0$ & yes & yes & no\\
Lattice run 8 & $3.9$ to $4.2$ & yes & yes & yes\\
Lattice run 12 & $5.1$ to $5.6$ & no & no & no\\
\addlinespace[2pt]
Runs 6 and 8 with tail & $1.0$ & yes & yes & yes\\
Run 12 with tail & $1.0$ & yes & yes & no\\
\bottomrule
\end{tabular}
\caption{Whether lithium solutions remain at the three test points of Sec.~\ref{sec:variations} when the emission model or the charge distribution changes, one change per row, and the resulting change of the peak lifetime $\tau_*$ at which $10^5\dh=2.508$.
Yes$^*$: within the statistical error of one simulation, about $5\%$ in $R_Q$ at the \yp point and $1\%$ at the other two; marginal: $^7$Li enters the stellar band just as \yp reaches its upper limit.
With tail: the lattice distribution continued above its last point by Eq.~\eqref{eq:S-distribution}.}
\label{tab:variations}
\end{table}

\begin{itemize}
\item \emph{Lifetime-charge relation.}
If the emission rate per unit charge depends on the energy of the emitted particles as $(\omega_Q/\omega_{\rm peak})^{\pm1}$, the lifetime grows with the charge as $\tau_Q=\tpeak x^{(5\pm1)/4}$ instead of $x^{5/4}$, and we test these two relations.
We implement them as a remapping of lifetime and charge that keeps the evaporation profile.
Each charge bin evolves as the soliton of Eq.~\eqref{eq:S-evolution} whose charge gives the new lifetime, so its mass loss keeps the form $(1-t/\tau_Q)^{3/5}$, and its particles start within $20\%$ of $\omega_Q(x,0)$ for $0.03\leq x\leq8$.
A modified emission rate would also change the exponent of the mass loss, to $3/(5\pm1)$.
The large solitons then evaporate earlier or later, which changes $\tau_*$ by $30$ to $40\%$, but the solutions remain at all three points within the statistical error, so the timing of the late emission shifts the lifetime without removing them.
\item \emph{Energy spread of the emitted particles.}
Since the emitted particles are created in pairs, their energies spread around $\bar\omega$ (Sec.~\ref{sec:source}).
We test a broad spread: each emitted particle has a spectrum proportional to $E$ between zero and $1.5\bar\omega$, with mean $\bar\omega$.
In calculations with $N_\nu=10^6$ and $\epsilon=-1$, the lithium solutions persist with this spread wherever they exist without it, for $\omega_{\rm peak}$ from $28.5$ to $60\,\MeV$ at $b=1$ and from $45$ to $80\,\MeV$ at $b\approx0.3$.
The more energetic antineutrinos convert more protons, which moves the $^7$Li edge to about $10\%$ smaller $R_Q$.
They also break up helium into more energetic fragments, which raise $^6\mathrm{Li}/{}^7\mathrm{Li}$ at the bounds by factors of $1.4$ to $1.9$; \neff changes by about $0.01$.
\item \emph{Muon channels.}
In the supersymmetric model of Sec.~\ref{sec:embedding}, $\mu^-\nu_e$ pairs replace a fraction $f_\mu$ of the $\nu_\mu e^-$ pairs once $\bar\omega>m_\mu/2$.
For $\epsilon=-1$, the excess antiquanta then emit $\bar\nu_e\mu^+$ in place of $\bar\nu_\mu e^+$: the direct $\bar\nu_e$ convert more protons, and the muon decays give most of their energy to neutrinos.
We compute populations 2 and 3 with these pairs (rows $\mu$ of Table~\ref{tab:soft-points}) and three more energetic populations, with $\omega_{\rm peak}=90$ and $120\,\MeV$, each in four simulations.
All five remain lithium solutions.
The pairs raise \neff by $0.005$ to $0.016$ and $^6\mathrm{Li}/{}^7\mathrm{Li}$ by up to $18\%$, and \yp still limits population 3 (rows $\mu$ of Table~\ref{tab:soft-points}).
The largest replacement allowed by electric neutrality, $f_\mu=\frac34\min(2S,1)$, with $S$ the muon suppression factor of Sec.~\ref{sec:embedding}, raises \neff by at most a further $0.003$ and lowers \li by at most a further $5\%$.
Table~\ref{tab:solutions} and the map of lithium solutions use the emission channels of Sec.~\ref{sec:source}, without these pairs.
\item \emph{Small solitons.}
Eq.~\eqref{eq:S-distribution} places a quarter of the energy in solitons below $x=0.1$.
They evaporate within about $10^3\,\mathrm{s}$ and emit the energetic antineutrinos that remove $^7$Be.
Cutting the distribution at $x=0.01$ or $0.04$ moves this energy to slightly larger solitons, which still evaporate early and lower \li further; the solutions remain at all three points.
Removing all solitons below $x=0.1$ raises \li at equal $R_Q$ by $13$ to $38\%$: the solutions survive at the \yp and $^7$Li points but disappear at the \neff point.
\item \emph{Large solitons.}
The Gaussian factor sets how much energy the largest solitons keep for the late cascades that destroy deuterium.
Replacing $e^{-x^2/4}$ by $e^{-x^2/2}$ or $e^{-x^2/8}$ changes $\tau_*$ by $+56$ or $-35\%$ and keeps the solutions.
A distribution growing as $x$, not as $x^{1/2}$, with only $6\%$ of the energy below $x=0.1$, removes them, except marginally at the $^7$Li point, where \li enters the stellar band just as \yp reaches its upper limit.
\item \emph{Lattice runs.}
The lattice simulations of Ref.~\cite{Kasuya:2025nix} scatter around Eq.~\eqref{eq:S-distribution} (Sec.~\ref{sec:source}).
We take $W$ directly from five of them, series 1, 2, 6, 8, and 12 of Fig.~4 there, normalized to the same $R_Q$.
Runs 1 and 2 hold almost half of the energy above $x=1$, whereas runs 6, 8, and 12 end near $x=1$.
Without these largest solitons, \dh reaches its measured value only at a peak lifetime four to six times longer.
Run 12 also has half the weight of Eq.~\eqref{eq:S-distribution} between $x=0.01$ and $0.1$, so it lacks part of the early antineutrinos and loses the solutions; at the \neff point, the longer lifetimes raise \neff, and most runs lose them there.
Continued above their last point by Eq.~\eqref{eq:S-distribution}, runs 6, 8, and 12 bring $\tau_*$ back within $2\%$ of the reference.
Table~\ref{tab:variations} gives the outcome at each test point.
\end{itemize}
Lithium solutions thus depend on the balance between the early antineutrinos and the late electromagnetic emission, with different sensitivities at the three test points.

\section{CMB spectral distortions from particles and Q-ball populations}\label{sec:cmb}
In this section, we show that the electromagnetic energy released by the particles and the Q-ball populations, and the later heating by residual energetic neutrinos, leave the CMB spectrum within the COBE/FIRAS bounds.

Photon production at high redshift thermalizes energy injected early~\cite{Khatri:2012tv,Chluba:2013kua}.
With the blackbody visibility $\mathcal J(z)\simeq\exp[-(z/z_{\rm dc})^{5/2}]$ and $z_{\rm dc}=1.98\times10^6$~\cite{Chluba:2013kua}, a small injection gives the chemical potential
\begin{equation}
 \mu\simeq1.401\int\mathcal J(z)\frac{dQ_{\rm em}}{\rho_\gamma},
 \label{eq:visibility}
\end{equation}
where $dQ_{\rm em}$ is the electromagnetic energy deposited per volume.
The COBE/FIRAS measurements bound $|\mu|<9\times10^{-5}$ and $|y|<1.5\times10^{-5}$~\cite{Fixsen:1996nj}.

The particles of Table~\ref{tab:realizations} decay exponentially with lifetimes near $9\times10^3\,\mathrm{s}$, and for energy fractions up to $10\%$, Eq.~\eqref{eq:visibility} gives $|\mu|<10^{-30}$.

A Q-ball population stops emitting when its largest solitons, with $x=8$, evaporate.
For the lithium solutions, with $\tpeak\leq2.6\times10^4\,\mathrm{s}$, this happens above $T_\gamma=1.9\,\keV$, at redshifts above $4z_{\rm dc}$, where $\mathcal J$ is negligible; the populations of Table~\ref{tab:soft-points} stop at $T_\gamma=2.4$ to $3.1\,\keV$, above $5z_{\rm dc}$.
The calculations behind Fig.~\ref{fig:region} extend to $\tpeak=10^7\,\mathrm{s}$, where the solitons would evaporate in the era of $\mu$ distortions, but no population with $\tpeak\gtrsim5\times10^4\,\mathrm{s}$ satisfies Eq.~\eqref{eq:S-abundances}.
The allowed populations stop emitting by $z\approx8\times10^6$, and along the upper edge of the allowed regions, with all released energy assigned to photons, Eq.~\eqref{eq:visibility} gives $|\mu|<10^{-24}$.

Residual energetic neutrinos keep heating the plasma after the transport calculation ends.
We bound this heating with the weak cross section $100G_F^2s/\pi$, giving the full incident energy to photons in every scattering.
During radiation domination, the transfer rate falls as $a^{-5}$ and integrates to its final value divided by $3H$.
The resulting $\mu$ and $y$ stay more than two orders of magnitude below these limits, and this energy is negligible for recombination~\cite{Slatyer:2016qyl}.

\section{Microscopic models of the Q-balls}\label{sec:embedding}
In this section, we discuss which microscopic models produce the Q-ball populations of the lithium solutions.

\subsection{What any microscopic model must provide}\label{sec:provide}
For a plateau of height $V_0$, Eq.~\eqref{eq:S-mass} holds with $C=\sqrt2\pi V_0^{1/4}$, and the field at the center of a Q-ball reaches $\phi_0=V_0^{1/4}Q^{1/4}$~\cite{Kusenko:1997si}.
The computed lithium solutions require $V_0^{1/4}$ from $32\,\mathrm{TeV}$, for population 3, up to $230\,\mathrm{TeV}$.
At the most probable charge, $\phi_0$ then lies between $2\times10^{11}$ and $2\times10^{12}\,\mathrm{GeV}$.

A scalar with Standard Model gauge charges gives the gauge bosons masses of order $g\phi$, and their loops lift the potential far above $V_0$ unless supersymmetry cancels them.
A condensate of sleptons must therefore lie along a supersymmetric flat direction, and it comes with sleptons below a few TeV and a light gravitino (Secs.~\ref{sec:susy} and \ref{sec:formation}).
A gauge singlet can instead take its plateau from a hidden sector~\cite{Heeck:2023idx,Kim:2024gmv} and turn into neutrino pairs through operators such as $\phi\phi(LH)(LH)/\Lambda^3$.
It then fixes no particle mass within collider reach, and its emission differs from that of an electron population.

A condensate with net lepton number breaks up into Q-balls above $T_{\rm sph}\simeq132\,\mathrm{GeV}$, the temperature below which sphalerons stop converting lepton number into baryon number~\cite{DOnofrio:2014rug}.
The solitons release their charge only long after, but lepton number left outside them turns partly into baryon number, with positive sign for $\epsilon<0$.
For population 3, this gives the observed baryon asymmetry if a fraction of about $3\times10^{-5}$ of the charge escapes capture.
Lattice simulations find more than $98\%$ of the charge in Q-balls, a bound set by their resolution~\cite{Kasuya:2000wx}.
Symmetric populations, $\epsilon=0$, carry no net lepton number and are free of this condition.

\subsection{A supersymmetric realization}\label{sec:susy}
An Affleck-Dine condensate~\cite{Affleck:1984fy,Dine:1995kz} along the flat direction $L_\mu L_eE_e^c$ consists of $\widetilde\nu_\mu$, $\widetilde e_L$, and $\widetilde E_e^c$ with equal amplitudes.
A small admixture of tau sleptons, of relative size $y_e/y_\tau$, cancels the F term of $H_d$ that the electron Yukawa coupling would otherwise leave.
The charge of the condensate, $q=3L+4Y$, with $L$ the lepton number and $Y$ the hypercharge, equals one on each field, so a Q-ball of charge $Q$ carries the lepton number $L_\mu=Q/3$.

At the soliton surface, pairs of quanta turn into pairs of electrons or neutrinos through gaugino exchange~\cite{Kawasaki:2012gk}.
The D terms keep the three fields equal, so the emission contains $\nu_\mu$, $e^-$, and $e^+$ in equal numbers, $b\simeq1$, as for population 3.
Emission of $\nu_e\bar\nu_e$ in place of $e^-e^+$ requires electroweak symmetry breaking, and we take it to stay subdominant, so populations 1, 2, and 4, with $b=0.75$, $0.9$, and $0.7$, need more neutral emission than this direction gives.
For $\omega_Q>m_\mu/2$, gaugino exchange also emits $\mu^-\nu_e$ pairs, whose rate relative to a massless pair is $S=(1-z_\mu)^3(1+3z_\mu)$ in saturated emission~\cite{Cohen:1986ct,Kawasaki:2012gk}, with $z_\mu=m_\mu/(2\omega_Q)$.
Since the Q-ball stays electrically neutral, they replace a fraction $f_\mu=3S/[2(1+2S)]$ of the $\nu_\mu e^-$ pairs, up to one half for $\omega_Q\gg m_\mu$; Sec.~\ref{sec:variations} gives their effect.

Gauge mediation links the plateau to the slepton masses.
Above the mass $M$ of $N$ pairs of messengers with F term $F$, it flattens the potential of the canonical field $\Phi$ of the direction into the plateau $V\simeq M_F^4\ln^2(|\Phi|^2/M^2)$, with $M_F$ set by $F$ and $N$~\cite{deGouvea:1997afu}, and $\zeta M_F=V_0^{1/4}$ with $\zeta\simeq5$~\cite{Kasuya:2025nix}.
The same messengers give the sleptons of the direction masses of order $F/M$, and since $F<M^2$, the lightest charged slepton weighs at most about $0.011\,\zeta M_F$~\cite{deGouvea:1997afu,Dimopoulos:1996gy,Martin:1996zb}.
This is $0.3$ to $0.4\,\mathrm{TeV}$ for population 3 and at most $3\,\mathrm{TeV}$ for the computed lithium solutions.

Searches for slepton pairs decaying into charged leptons and gravitinos~\cite{ATLAS:2019lff,CMS:2020bfa}, and for long-lived charged particles~\cite{ATLAS:2019gqq}, exclude most of these masses for population 3 when a charged slepton is the lightest superpartner of the Standard Model fields.
Population 3 remains allowed when sneutrinos are the next-to-lightest superpartners, below the charged sleptons and above the gravitino, as general gauge mediation permits~\cite{Meade:2008wd,Katz:2009qx}.
The charged sleptons then decay into the sneutrinos with soft charged leptons, which searches for compressed sleptons do not yet reach~\cite{ATLAS:2025evx}.

With the Higgs sector of the minimal supersymmetric Standard Model, the stops give $m_h\leq116\,\mathrm{GeV}$ for population 3~\cite{Allanach:2001kg}.
Couplings of the Higgs doublets to the messengers~\cite{Craig:2012xp}, vector-like matter~\cite{Martin:2009bg,Endo:2011mc,Evans:2011bea}, or a singlet can raise $m_h$ to its measured value; both doublets vanish on the flat direction, so these couplings leave its potential unchanged at tree level.

\subsection{Formation of the Q-balls}\label{sec:formation}
The plasma created by inflaton decay adds a thermal logarithm to the potential~\cite{Anisimov:2000wx,Kasuya:2003yr}, which holds the condensate at large field values until the temperature falls below about $10^5\,\mathrm{GeV}$~\cite{Kasuya:2003yr}.
The Q-balls then form in one of two ways.

If the reheating temperature exceeds a few TeV, the condensate is released while the inflaton still dominates, after its phase has settled at a minimum of the A term.
It then oscillates almost along a line and fragments into Q-balls of both signs with comparable charges~\cite{Kasuya:2001hw,Kasuya:2003yr}, which form symmetric populations.

All supersymmetry breaking then comes from the messengers, and the gravitino weighs about $0.5$ to $10\,\mathrm{eV}$ for $\omega_{\rm peak}$ from $28.5$ to $60\,\MeV$.
The CMB, weak lensing, and the galaxy power spectrum exclude such a relic gravitino above $1$ to $2\,\mathrm{eV}$~\cite{Xu:2021rwg}, unless entropy released after its decoupling dilutes it by up to a factor of three to four.
Its contribution to \neff then falls from $0.06$ to about $0.01$.
These symmetric populations, with $b=1$ and the $\mu^-\nu_e$ pairs, remain lithium solutions at $\omega_{\rm peak}=45$ and $60\,\MeV$, with \neff near $3.0$; at $28.5$ and $34\,\MeV$, \yp reaches its upper limit before $^7$Li enters the stellar band.

If supersymmetry breaking has a total F term much larger than that of the messengers, gravity mediation gives the direction the mass $m_{3/2}$.
The gravity-mediated quadratic term dominates the plateau above the field value $\phi_{\rm eq}\sim M_F^2/m_{3/2}$.
The condensate then starts to oscillate at $H\simeq m_{3/2}$ on an orbit of order-one ellipticity, and it fragments into Q-balls of one sign, $\epsilon=\pm1$, once its amplitude falls to $\phi_{\rm eq}$~\cite{Kasuya:2001hw,Kawasaki:2002hq,Akita:2025zvq}.
Since $\phi_{\rm eq}$ fixes both $Q_{\rm peak}$ and $\omega_{\rm peak}$, population 3 requires $m_{3/2}=0.5$ to $2\,\MeV$, and its abundance follows from reheating temperatures of $10$ to $300\,\mathrm{GeV}$.
Q-balls of gravity mediation would need hundreds of oscillations to form~\cite{Hiramatsu:2010dx}, longer than the condensate oscillates before the plateau takes over, and the gravitinos made at these temperatures stay below the dark matter density.

In both histories, the solitons form at $10^4$ to $10^5\,\mathrm{GeV}$, above $T_{\rm sph}$, and keep nearly all their charge in the cooling plasma~\cite{Kasuya:2001hw,Kawasaki:2002hq}.
Their largest charges scale with the fourth power of the initial amplitude, as at zero temperature~\cite{Kasuya:2010vq}, so Eq.~\eqref{eq:S-distribution} applies if fragmentation reproduces the distribution found in lattice simulations~\cite{Kasuya:2025nix}.
Under this condition, and if the Q-balls capture nearly all of the charge, the supersymmetric model gives the lithium solutions with $b\simeq1$, such as population 3, with an MeV gravitino, an additional contribution to $m_h$, and sneutrinos as the next-to-lightest superpartners.
In this model, sleptons below a few TeV accompany every lithium solution.

\end{document}